\documentclass[useAMS,usenatbib,a4paper]{mn2e}

\usepackage{float}
\usepackage{bm}
\usepackage{epsfig}
\usepackage{soul}
\usepackage{subcaption}
\usepackage{multirow,balance}
\usepackage{amssymb}
\usepackage{amsfonts}
\usepackage{amsmath}
\usepackage{aas_macros}
\usepackage{graphicx}
\usepackage{natbib}
\usepackage{color}
\usepackage{listings}
\usepackage{hyperref}  
\hypersetup{dvips, colorlinks=true, linkcolor=black, citecolor=black, filecolor=black, urlcolor=black}

\definecolor{grey}{rgb}{0.75,0.75,0.75}
\definecolor{Orange}{rgb}{1.0,0.5,0.15}
\definecolor{brown}{rgb}{0.7,0.25,0.0}
\definecolor{pink}{rgb}{1.0,0.5,0.5}
\definecolor{darkerred}{rgb}{0.8,0,0}
\definecolor{darkerblue}{rgb}{0,0,0.8}
\definecolor{Blue}{rgb}{0,0.08,0.65}
\definecolor{Red}{rgb}{0.65,0.08,0.05}
\definecolor{Green}{rgb}{0.15,0.45,0.25}

\begin{document}

\setcounter{tocdepth}{3}

\title[Secular resonant orbital diffusion]
{Secular resonant dressed orbital diffusion II :    \\
application to  an isolated self similar tepid galactic disc 
}

\author[Jean-Baptiste Fouvry, Christophe Pichon]{Jean-Baptiste Fouvry$^{1,2}$ and Christophe Pichon$^{1,3}$
\vspace*{6pt}\\
\noindent$^{1}$ Institut d'Astrophysique de Paris, CNRS (UMR-7095), 98 bis Boulevard Arago, 75014, Paris, France\\
$^{2}$ UPMC Univ. Paris 06, UMR-7095, 98 bis Boulevard Arago, 75014, Paris, France\\
$^{3}$ Institute of Astronomy, University of Cambridge, Madingley Road, Cambridge, CB3 0HA, United Kingdom\\
}

\date{\today}
\label{firstpage}
\pagerange{\pageref{firstpage}--\pageref{lastpage}}

\maketitle

\begin{abstract}
The main orbital signatures of the secular evolution of an isolated self-gravitating stellar Mestel disc are recovered using 
a dressed Fokker-Planck formalism in angle-action variables. The shot-noise-driven formation of narrow ridges of resonant orbits is recovered 
in the WKB limit of tightly wound transient spirals, for a tepid Toomre-stable tapered disc.
The relative effect of the bulge, the halo, the disc temperature and the spectral properties of the shot noise are investigated in turn.
For  such galactic discs all elements seem to impact the locus and 
 direction of the ridge.
For instance, when the halo mass is decreased, 
we observe a transition 
 between a regime of heating in the inner regions of the disc through the inner Lindblad resonance to a regime of radial migration of quasi-circular orbits via the corotation resonance in the outer part of the disc.  
 The dressed secular formalism captures both the 
nature of  collisionless systems (via their natural frequencies and susceptibility), and their nurture via the structure of the external perturbing power spectrum.
Hence it provides the ideal framework in which to study their long term evolution. 
\end{abstract}

\begin{keywords}
Galaxies, dynamics, evolution, diffusion
\end{keywords}

\section{Introduction}
\label{sec:introduction}

Understanding the dynamical secular evolution of galactic discs over cosmic times is a long-standing endeavour, which recent renewed interest as their
cosmological environment  is now more firmly established in the context of the $\Lambda$CDM paradigm \citep{Planck2013Cosmoparams}. 
Indeed, disentangling the respective role of the cosmic environment (nurture) and the intrinsic structure of galaxies (nature)
in explaining the observed physical and morphological distribution of galaxies
 is  focussing much recent activities. 
Self-gravitating discs are cold  dynamical systems, for which rotation represents an important reservoir of free energy. Some perturbations are strongly amplified, while resonances tend to confine and localise their dissipation: even small stimuli can lead to discs evolving to distinct equilibria.
 
 Modern N-body simulations now allow for 
both detailed modeling of intricate non-linear physical processes \citep[such as multi-scale hydrodynamics, star formation, AGN feedback... see e.g.][]{Dubois2014}, but also well controlled idealized numerical experiments \citep{SellwoodAthanassoula1986,Earn1995,Sellwood2012}.
Such experiments are essential to understand how the orbital structure of a galactic disc may drive its secular evolution and simply take into account 
their self gravity.
Yet the reliability of numerical simulations to preserve the symplectic nature of the underlying physical system over hundreds of orbital times 
is  potentially an issue which calls for alternative probes.

 In parallel, over the past few years it has been found that the formalism of angle-action variables \citep{born1960mechanics,Goldstein} and the so-called matrix method \citep{Kalnajs2}  also
provided means of accounting for the self-gravity in the secular equation driving such systems in the limit of large number of particles.
They are in particular well suited to disentangling the intricate roles of resonances and identifying the orbital families driving 
their secular evolution.

 Two equations for secular diffusion have 
recently been revisited  using these coordinates: 
i) the (possibly dressed) Fokker-Planck equation (\citealt{Binney1988}; \citealt{weinberg93}; \mbox{\citealt{Pichon2006}}; \citealt{Chavanis2012EPJ}; \citealt{Nardini2012};  the companion paper, hereafter paper I, \citealt[][submitted]{FouvryPichonPrunet2014}), when the 
source of (possibly coloured) potential fluctuations is taken to be an external bath, e.g. corresponding to the cosmic environment;
ii) the Balescu-Lenard non-linear equation \citep{Balescu1960,Lenard1960,Weinberg1998,weinberg2001b,Heyvaerts2010,Chavanis2012} which accounts for  self-driven orbital
secular diffusion induced by  shot noise corresponding to the discreteness of the system.
These equations are fairly unique in providing a theoretical framework for the secular evolution of stellar and dark matter dominated systems,
and  well suited to explain, complement and understand the results of these crafted numerical experiments. 
As an illustration of their versatility 
we will implement here the (simpler) Fokker-Planck equation in order to \textit{explain} one such experiment, and postpone the implementation of the Balescu-Lenard equation to further investigation.

Indeed, \cite{Sellwood2012} (hereafter S12) has suggested, using a well controlled numerical setting, that an isolated stable stellar disc would secularly drift towards a state of marginal stability through the spontaneous generation of transient spiral structures. In its experiment, S12 evolves a set of particles, of increasing number, for hundreds of dynamical times. These particles are distributed according to equation~\eqref{DF_Jr_Jphi_isothermal} below which corresponds to what should be a stable distribution. Nonetheless, S12 identifies the noise driven formation of \textit{ridges} in action space, along very specific resonant directions.
Indeed, discs which are fully stable at a linear level may still on the long-term develop spiral structure that can grow to important amplitudes, 
eventually transforming the disc into a barred-type spiral galaxy.

We intend to show here how the formalism of \textit{dressed}  secular diffusion written in angle-action variables
 is able to predict  the observed  drifting process  exhibited in S12.  
 The direct analytical or numerical calculation of the modes of a galactic disc is a complex task, which has only been made for a small number of disc models \citep{Zang1976,Kalnajs1977,Vauterin1996,Pichon1997,Evans1998,Jalali2005}.
For the sake of analytical simplicity, following paper I, 
 we will  make use of the so-called WKB approximation (\citealt{WKB}; \citealt{Toomre1964}; \citealt{Kalnajs1965}; \citealt{Lin1966}), and assume that the initial distribution is 
 well described by a tepid Schwarzschild distribution function, which will allow us to compute the gravitational susceptibility of 
 our disc as a simple (scalar) multiplicative factor,
 and express the diffusion coefficient  algebraically.  This approximation provides insight into the location of the relevant resonances. 
 Our strategy here is to
defer to appendices  as much of the technical detail as we can while still
conveying an understanding of the overall theory in the main text (see also paper I).
The main orbital signature of the secular evolution of the  tepid Toomre-stable tapered Mestel disc will be recovered   
in the WKB limit of tightly wound transient spirals. The relative effect of the bulge, the halo, the disc temperature and the spectral properties of the shot noise will be  investigated in turn.

  The paper is organized as follows.  Section \ref{sec:model} introduces briefly the   \cite{Mestel1963} disc considered by  S12 and presents its main features. Section \ref{sec:seculardiffusion} recaps the formalism of the secular diffusion equation in action-space, which  is able to \textit{capture} the main observed features  when considered in the WKB limit for a tepid disc.
Our results for the Mestel disc are presented in section \ref{sec:results}. Finally, we conclude in section \ref{sec:conclusions}.
Appendices~\ref{sec:derivation_diffusion_equation} and \ref{sec:WKB_approximation} present respectively a rapid derivation of the diffusion equation and its WKB limit; paper I presents a more systematic derivation.

\section{The disc model}
\label{sec:model}

\begin{figure}
\begin{center}
\epsfig{file=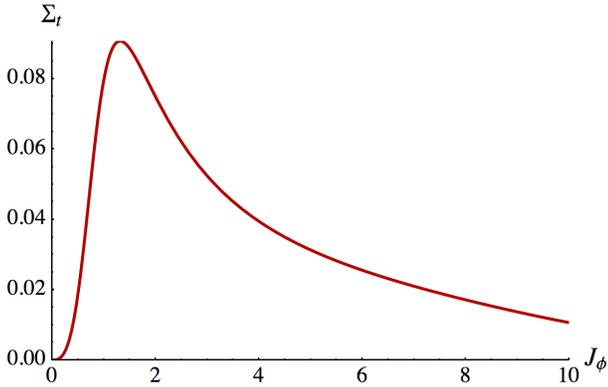,angle=-00,width=0.45\textwidth}
\caption{\small{Surface density $\Sigma_{\rm t}$ of the tapered Mestel disc. The unit system has been chosen so that $V_{0} = G = R_{i} = 1$.
}}
\label{figSigmaDF}
\end{center}
\end{figure}

\begin{figure}
\begin{center}
\epsfig{file=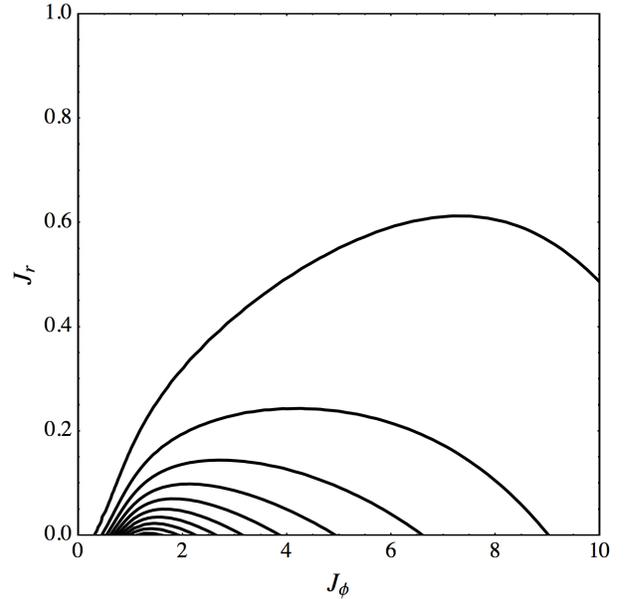,angle=-00,width=0.45\textwidth}
\caption{\small{Contours of the initial distribution function in action-space $(J_{\phi},J_{r})$, within the epicyclic approximation. The contours are spaced linearly between 95\% and 5\% of the distribution function maximum.
}}
\label{figcontourDF}
\end{center}
\end{figure}

\begin{figure}
\begin{center}
\epsfig{file=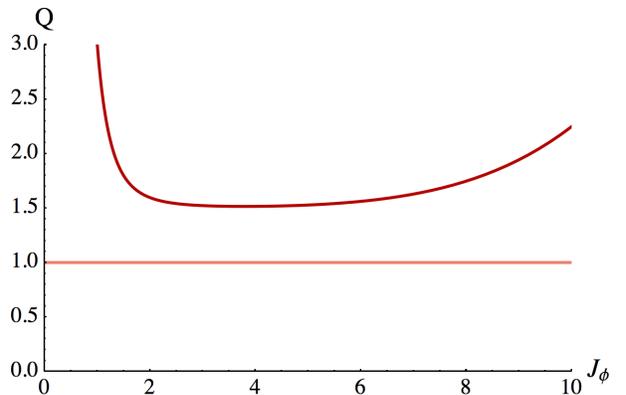,angle=-00,width=0.45\textwidth}
\caption{\small{Variation of the local $Q$ Toomre parameter with the angular momentum $J_{\phi}$. It is scale invariant except in the inner/outer regions because of the presence of the tapering functions $T_{\rm inner}$ and $T_{\rm outer}$. The unit system has been chosen so that ${ V_{0} = G = R_{i} = 1 }$.
}}
\label{figQDF}
\end{center}
\end{figure}

Stellar discs are dynamical systems, which at leading order have reached a virialized state within an axisymmetric gravitational field created not only by their own mass, but also by other constituents of the galaxy, mainly the inner bulge and the surrounding dark-matter halo.
The disc considered by  S12 is an infinitely thin Mestel disc \citep{Mestel1963}, for which the circular speed ${v_{\phi}^{2} = R \, {\partial \psi_{0}}/{\partial R} = V_{0}^2}$ of the stars is independent of the radius. The stationary potential background of such a disc and its associated surface density are given by
\begin{equation}
\psi_{0} (R) = V_{0}^{2} \ln \!\left[ \frac{R}{R_{i}} \right] \;\;\; ; \;\;\; \Sigma (R) = \frac{V_{0}^{2}}{2 \pi G R} \, ,
\label{potential_surface_density_Sellwood}
\end{equation}
where $V_{0}$ and $R_{i}$ are scale parameters. Let us introduce the actions of the system,  $(J_{r} , J_{\phi})$ \citep{born1960mechanics,BinneyTremaine2008}. For an axisymmetric two dimensional disc, the azimuthal action $J_{\phi} = L_{z}$ is the angular momentum, whereas the radial action is given by ${J_{r} = 1/ \pi \int_{R_{\rm min}}^{R_{\rm max}} v_{R} \, \mathrm{d} R }$ and encodes the amount of radial energy of a star, where ${ v_{R} \!=\! [ 2 (E \!-\! \psi_{0} (R)) \!-\! J_{\phi}^{2} / R^{2} ]^{1/2} }$, is integrated between the pericenter $R_{\rm min}$ and apocenter $R_{\rm max}$ of the trajectory. Here $J_{r} = 0$ corresponds to circular orbits.
These action remain well defined through the secular evolution of the disc as cylindrical symmetry is preserved. For simplicity we use the epicyclic approximation to describe the behavior of the distribution function of the system in the action-space $(J_{r},J_{\phi})$. This approximation holds as long as the particles do not have too eccentric orbits. Because the Mestel disc has constant circular velocities, the link between the angular momentum $J_{\phi}$ and the guiding radius $R_{g}$ is straightforward and reads
\begin{equation}
J_{\phi} = R_{g} \, V_{0} \,.
\label{link_Jphi_Rg}
\end{equation}  
Within the epicyclic approximation, one can obtain the expression of the \textit{intrinsic} frequencies, the azimuthal frequency $\Omega (J_{\phi})$ and the epicyclic frequency $\kappa (J_{\phi})$, given by
\begin{equation}
\begin{cases}
\begin{aligned}
\displaystyle \Omega (J_{\phi}) &= \bigg[ \frac{1}{R_{g}} \frac{\partial \psi_{0}}{\partial R} \bigg|_{R_{g}} \bigg]^{1/2} \!\!\!\! = \frac{V_{0}^{2}}{J_{\phi}}\,,
\\
\displaystyle \kappa (J_{\phi}) &= \bigg[ \frac{\partial^{2} \psi_{0}}{\partial R^{2}} \bigg|_{R_{g}} \!\!\!\!\!+ 3 \frac{J_{\phi}^{2}}{R_{g}^{4}} \bigg]^{1/2} \!\!\!\!= \sqrt{2} \, \Omega (J_{\phi}) \, .
\end{aligned}
\end{cases}
\label{expression_intrinsic_frequencies}
\end{equation}
Two remarks should be made on these frequencies. First, within the epicyclic approximation, the two frequencies are only function of the angular momentum $J_{\phi}$ and do not depend on the radial action $J_{r}$. Moreover, one should note that we have ${ {\kappa}/{\Omega} = \sqrt{2} }$, so that the Mestel disc is an intermediate case between the Keplerian case for which ${ {\kappa}/{\Omega} = 1 }$, and the harmonic case for which ${ {\kappa}/{\Omega} = 2 }$. The ratio between the intrinsic frequencies is an important parameter for the dynamical behavior of the system, since it determines the location of the resonances and a constant ratio may introduce degeneracies. Using the epicyclic approximation, the distribution function considered by S12, takes the form of a locally \textit{isothermal}-DF or Schwarschild-DF, which reads
\begin{equation}
F_{0} (J_{r} , J_{\phi}) = \frac{\Omega (J_{\phi}) \, \Sigma_{\rm t} (J_{\phi}) }{\pi \, \kappa (J_{\phi}) \, \sigma_{r}^{2}} \, \exp \left[ - \frac{\kappa (J_{\phi})}{\sigma_{r}^{2}} J_{r} \right] \,,
\label{DF_Jr_Jphi_isothermal}
\end{equation}
where the intrinsic frequencies are given by equation~\eqref{expression_intrinsic_frequencies}, $\sigma_{r}$ is a constant dispersion describing the spread in radial velocity of the distribution function,  and the taped surface density $\Sigma_{\rm t}$ in analogy with equation~\eqref{potential_surface_density_Sellwood} is given by
\begin{equation}
\Sigma_{\rm t} (J_{\phi}) = \frac{V_{0}^{3}}{2 \pi G  J_{\phi}} \, T_{\rm inner} (J_{\phi}) \, T_{\rm outer} (J_{\phi}) \,.
\label{Sigma_taped}
\end{equation}
In equation~\eqref{Sigma_taped}, $T_{\rm inner}$ and $T_{\rm outer}$ are tapering functions used to damp out the contributions from the inner and outer regions, which read
\begin{equation}
\begin{cases}
\displaystyle T_{\rm inner} (J_{\phi}) = \frac{J_{\phi}^{\nu}}{(R_{i} V_{0})^{\nu} + J_{\phi}^{\nu}} \,,
\\
\displaystyle T_{\rm outer} (J_{\phi}) = \left[ 1 + \left[ \frac{J_{\phi}}{R_{0} V_{0}} \right]^{\mu} \right]^{-1} \!\!.
\end{cases}
\label{inner_outer_taper_Sellwood}
\end{equation}
where $\nu$ and $\mu$ control the sharpness of the two tapers. The inner tapering function at the scale $R_{i}$ induces an important distribution function gradient at this scale, which is indeed responsible for the position of the peak of diffusion and reflects the presence of a bulge. 
The outer tapering function reflects the finite size of the disc.
For the numerical simulations, we used the same constants as in S12. We place ourselves in a unit system such that: ${ V_0 = G = R_{i} = 1 }$. The other numerical factors are given by ${ \sigma_{r} = 0.284 }$, ${ \nu = 4 }$, ${ \mu = 5 }$, ${ R_{0} = 11.5 }$.
The shape of the damped surface density $\Sigma_{\rm t}$ is shown on figure~\ref{figSigmaDF}.
The tapering functions have for effect to reduce and turn off the self-gravity of the disc in the inner and outer regions. 
As such, it provides a fairly general class of models for more realistic discs.
One can now look at the initial level contours of the distribution function represented on figure~\ref{figcontourDF}.
One of the consequences of such a scale-invariant disc is that its local Toomre Parameter $Q$ \citep{Toomre1964} is almost independent of the radius for the intermediate regions. Indeed, one defines $Q$ as
\begin{equation}
Q = \frac{\sigma_{r} \, \kappa (J_{\phi})}{3.36 \, G \, \xi \, \Sigma_{\rm t} (J_{\phi})} \,,
\label{definition_Q}
\end{equation}
where in order to reduce the \textit{susceptibility} of the disc, we suppose that only a fraction, $\xi$, of the disc is self-gravitating, with $0 \leq \xi \leq 1$, so that the 
rest of the gravitational field is provided by the halo. For S12's simulation, the fraction of active surface density was $\xi = 0.5$. The dependence of $Q$ with  radius is represented on figure~\ref{figQDF}. The scale invariance of this stability parameter leads as expected to a constant $Q \simeq 1.5$ throughout most of the disc. It is only broken by the presence of the tapering functions which damp the surface density $\Sigma_{\rm t}$ for the most inner and outer regions.

\section{The secular diffusion equation}
\label{sec:seculardiffusion}

The figure $7$ of  S12  exhibits  a ridge best seen in the contours of the distribution function in  $(J_{\phi},J_{r})-$space. This irreversible diffusion feature of the DF was obtained through transient spiral features, while evolving a stationary Mestel disc sampled by pointwise particles. The formalism of  secular diffusion should allow us to  predict and explain its appearance. Let us first recall the main equations governing secular diffusion. The secular dynamics intends to describe the long-term aperiodic evolution of a self-gravitating system, during which small resonant and cumulative effects can add up in a coherent way. These small effects, amplified through the self-gravity of the system, can be seeded in two manners.
 As argued earlier, a possible  first approach to describe such seeds
 is the secular diffusion equation 
 which describes the long-term evolution of a collisionless self-gravitating system undergoing \textit{external perturbations}. The second approach is the Balescu-Lenard equation for inhomogeneous system 
  which describes the evolution of \textit{closed} self-gravitating system undergoing perturbations arising from its own discreteness. In the first case, the perturbations are exterior and the system responds to it, whereas in the second case the perturbations are intrinsic and self-induced.
Real discs respond to stimuli corresponding to a combination of these two formalisms. The finite number of stars in the disc, the presence of giant molecular clouds and of massive sub-halos around the disc all induce Poisson shot noise. Spiral arms in the gas distribution constitute another source of gravitational noise. Finally, the presence of a bar drives an additional coherent perturbation. The complex dynamical history of a real stellar disc encompasses responses to all these stimuli.
  
  The context of the evolution of the Mestel disc studied numerically by  S12 corresponds to the Balescu-Lenard formalism. The perturbations originates both from the discrete sampling of the smooth DF, which is only represented by a finite number of particles, but also from numerical noise which can induce long-term and irreversible diffusion. The formalism of the Balescu-Lenard equation is much more involved than that of the external secular diffusion
and will be the topic of future work. One can still however extract approximate yet interesting qualitative information for the long-term dynamics from the approach relying on external perturbation. In order to take into account the fact that the perturbations undergone by the system are created by the galactic disc itself, we will assume, as detailled later on, that the amount of noise a given location is generically proportional to the square-root of the local surface density, assuming that these intrinsic perturbations behave like a Poisson shot noise. 

\subsection{The dressed secular equation in action space}
\label{sec:diffusionequation}

The distribution function introduced in equation~\eqref{DF_Jr_Jphi_isothermal} is a stationary distribution function, since it depends only on the actions coordinates ${\bm{J} = (J_{\phi} , J_{r})}$. The long-term evolution of the distribution in action-space is given by the secular diffusion equation, derived briefly in Appendix~\ref{sec:derivation_diffusion_equation} (see also paper I for details). 
 which takes the form
\begin{equation}
\frac{\partial F_0}{\partial t} = \sum\limits_{\bm{m}} \bm{m} \!\cdot\! \frac{\partial }{\partial \bm{J}} \left[ D_{\bm{m}} (\bm{J}) \, \bm{m} \!\cdot\! \frac{\partial F_0}{\partial \bm{J}} \right] \,,
\label{diffusion_equation}
\end{equation}
where the diffusion coefficients $D_{\bm{m}} (\bm{J})$ are given by
\begin{equation}
\begin{aligned} \!\!\!\!\!
D_{\bm{m}} (\bm{J}) =  \frac{1}{2}\sum\limits_{p , q} \psi^{(p)}_{\bm{m}} \, \psi^{(q) *}_{\bm{m}} 
 \left[ [\mathbf{I} \!-\!  \widehat{\mathbf{M}}]^{-1} \!\cdot \widehat{\mathbf{C}}\!\cdot\![\mathbf{I} \!-\! \widehat{\mathbf{M}}]^{-1} \!\right]_{p q} \!\!\! .
\end{aligned}
\label{expression_diffusion_coefficients_all}
\end{equation}
In this expression the response matrix, $ \widehat{\mathbf{M}}$,  and the cross-power spectra $
\widehat{\mathbf{C}}$ (see below) are functions of  $\omega$ which  should be evaluated at the resonant frequencies $\bm{m} \!\cdot\! \bm{\Omega}$, where
 the index ${\bm{m} = (m_{r} , m_{\phi}) \in \mathbb{Z}^{2}}$ corresponds to the Fourier coefficients associated to the Fourier transform with respect to the angles $(\theta_{r} , \theta_{\phi})$ of the actions $(J_{r} , J_{\phi})$. To one specific $\bm{m}$ is associated one specific resonance. Throughout our calculation, we will restrict ourselves to only three different resonances which are: the inner Lindblad resonance (ILR) corresponding to ${(m_{r}^{\rm ILR} , m_{\phi}^{\rm ILR}) = (-1,2)}$, the outer Lindblad resonance (OLR) for which ${(m_{r}^{\rm OLR} , m_{\phi}^{\rm OLR}) = (1,2)}$, and finally the corotation resonance (COR) associated with circular motion which satisfies ${(m_{r}^{\rm COR}, m_{\phi}^{\rm COR}) = (0,2)}$. Indeed, S12 restricted disturbing forces to ${m_{\phi} \!=\! 2}$, so that we may impose the same restriction on the considered azimuthal number $m_{\phi}$. Moreover, all the estimations presented in the upcoming sections have also been performed while considering resonances with ${ m_{r} \!=\! \pm 2 }$, which were checked to be subdmoninant, so as to justify our resonances restriction to the ILR, COR and OLR. Equation~\eqref{expression_diffusion_coefficients_all} for the diffusion coefficients also involves potential basis elements given by $\psi^{(p)}$, as introduced in the Kalnajs matrix method \citep{Kalnajs2}. Here $\psi^{(p)}_{\bm{m}} (\bm{J})$ corresponds to the Fourier transform  of index $\bm{m}$ with respect to the angles $\bm{\theta}$. Another key element of equation~\eqref{expression_diffusion_coefficients_all} is the response matrix $\widehat{\mathbf{M}}$, which indicates how the system amplifies a given perturbation. It is given by
\begin{equation}
\!\!\!\widehat{\mathbf{M}}_{p q} (\omega) \!=\!  (2 \pi)^{2} \!\sum\limits_{\bm{m}} \!\!\int\!\! \mathrm{d}^{2} \!\bm{J} \frac{\displaystyle \bm{m} \!\cdot\! \partial F_{0} \!/\! \partial \bm{J}}{\omega \!-\! \bm{m} \!\cdot\! \bm{\Omega}} \!\left[  \psi^{(p)}_{\bm{m}} (\bm{J}) \!\right]^{*} \!\!\psi^{(q)}_{\bm{m}} (\bm{J}) \,,
\label{Fourier_M}
\end{equation}
where one can see that the pole at the intrinsic frequency ${\omega = \bm{m} \!\cdot\! \bm{\Omega}}$ plays a crucial role for the amplification.

In order to underline the physical meaning of these diffusion coefficients, one can rewrite them under the shortened  form
\begin{equation}
D_{\bm{m}} (\bm{J})  \sim \frac{\left< \left| \psi^{\rm ext}_{\bm{m}} (\omega) \right|^{2}\right>}{\left| \varepsilon (\bm{m},\omega) \right|^{2}} (\omega = \bm{m} \!\cdot\! \bm{\Omega}) \,,
\label{expression_diffusion_coefficients_shortened}
\end{equation}
where qualitatively (see \cite{Chavanis2012EPJ} for the homogeneous case), we have the following scalings 
\begin{equation}
\begin{cases}
\displaystyle \left< \left| \psi^{\rm ext}_{\bm{m}} (\omega) \right|^{2}\right>  \sim  \widehat{\mathbf{C}}\sim \left< \widehat{\bm{b}} \!\cdot\! \widehat{\bm{b}} ^{* \, t}\right> \, ,
\\
\displaystyle \frac{1}{\left| \varepsilon ( \bm{m} , \omega) \right|^{2}} \sim \left(\! \left[ \mathbf{I} \!-\! \widehat{\mathbf{M}} \right]^{-1} \!\right)^{\!2} \, ,
\end{cases}
\label{substitution_shortened_diffusion_coefficients}
\end{equation}
where the coefficients $\widehat{\bm{b}}$ correspond to the basis decomposition of the exterior perturbation so that ${\psi^{\rm ext} \!=\! \sum_{p} \!b_{p} (t) \psi^{(p)}}$. These coefficients are then Fourier transformed with respect to time and one only needs to study their statistical ensemble-averaged cross-correlation  defined in detail in equation~\eqref{definition_autocorrelation_exterior}. The diffusion coefficients are given by the ratio of the power spectrum of the external perturbations ${\langle \left| \psi^{\rm ext}_{\bm{m}} (\omega) \right|^{2} \rangle}$ divided by the gravitational susceptibility ${\left| \varepsilon (\bm{m},\omega) \right|^{2}}$ of the disc. 
We will suppose that the \textit{exterior} perturbation, which represents the intrinsic noise of the system has a particular spectrum, since it originates from the galactic disc itself, so that
\begin{equation}
\left< \left| \psi^{\rm ext} (J_{\phi} , \omega) \right|^{2}\right> \simeq \Sigma_{\rm t} (J_{\phi}) \,.
\label{assumption_noise}
\end{equation}
This assumption on the perturbations is relatively crude since we have only included a spatial dependence of the noise with $J_{\phi}$. For a system perturbed by a more realistic exterior source, the spectrum of perturbations is more \textit{coloured} and depends on the full statistical properties of the exterior perturbers. Here the lack of dependence with the temporal frequency $\omega$ implies that the three resonances ILR, OLR and COR undergo the same perturbations at each position, even if they are not associated to the same frequencies of resonances $\bm{m} \!\cdot\! \bm{\Omega}$. As a consequence, there is no distinctions between the perturbations \textit{felt} at the inner and outer Lindblad resonances. From the shape of the surface density in figure~\eqref{figSigmaDF}, one may see that the region of the inner tapering $(J_{\phi} \simeq 1.5)$ will be the most perturbed. With this assumption,  expression~\eqref{expression_diffusion_coefficients_all} for the diffusion coefficients may be rewritten under the much simpler form
\begin{equation}
D_{\bm{m}} (\bm{J}) \sim \frac{\Sigma_{\rm t} (J_{\phi}) }{\left| \varepsilon (\bm{m},\bm{m} \!\cdot\! \bm{\Omega}) \right|^{2}} \,.
\label{expression_diffusion_coefficients_simplified}
\end{equation}
Equation~\eqref{expression_diffusion_coefficients_simplified} implies that the secular response of the system is the result of an arbitration between the system intrinsic noise and its local susceptibility. 

The diffusion  equation~\eqref{diffusion_equation}  can be rewritten as the divergence of a flux in order to underline the exact conservation of total mass as
\begin{equation}
\frac{\partial F_{0}}{\partial t} = \sum\limits_{\bm{m}} \frac{\partial }{\partial \bm{J}} \!\cdot\! \left[ \bm{m} \, D_{\bm{m}} (\bm{J}) \,\!\left( \bm{m} \!\cdot\! \frac{\partial F_{0}}{\partial \bm{J}} \right) \right] \,.
\label{diffusion_equation_divergence}
\end{equation}
We define as $M (t)$ the mass contained in a volume $\mathcal{V}$ of the action-space at time $t$, so that
\begin{equation}
M (t) = \int_{\mathcal{V}} \! \mathrm{d}^{2} \!\bm{J} \, F_{0} (\bm{J} , t) \,.
\label{definition_mass_interpretation}
\end{equation}
Using the divergence theorem, the variation of mass due to secular diffusion can be seen as a flux of particles through the boundary $\mathcal{S}$ of this volume, with $\bm{n}$ being the corresponding exterior pointing normal vector, so that 
\begin{equation}
\frac{\mathrm{d} M}{\mathrm{d} t} = \sum\limits_{\bm{m}} \int_{\mathcal{S}} \! \mathrm{d} S \, (\bm{m} \!\cdot\! \bm{n}) \, D_{\bm{m}} (\bm{J}) \, \bm{m} \!\cdot\! \frac{\partial F_{0}}{\partial \bm{J}} \,.
\label{variation_mass_interpretation_II}
\end{equation}
In equation~\eqref{variation_mass_interpretation_II}, the contribution from a given resonance $\bm{m}$ corresponds to a preferential diffusion in the direction $\bm{m}$. This diffusion is anistropic in the sense that it is maximum for ${\bm{n} \propto \pm \bm{m}}$, and equal to $0$, for ${\bm{n} \!\cdot\! \bm{m} = 0}$. Two quantities influence the \textit{strength} of the diffusion. On the one hand, the diffusion coefficient $D_{\bm{m}} (\bm{J})$ describes how sensitive the system is at a given location in action-space. On the other hand, the gradient $\bm{m} \cdot \partial F_{0} / \partial \bm{J}$ describes how structured and inhomogeneous the system is at the same location. Such a formulation is of interest. First of all, it allows us to identify in each position $(J_{\phi},J_{r})$ what is the dominant resonance. It also permits us to identify the position of the peak of maximum flux, where the total flux, $\bm{\mathcal{F}}_{\rm tot}$, is defined as
\begin{equation}
\bm{\mathcal{F}}_{\rm tot} = \sum\limits_{\bm{m}} \, \bm{m} \, \!\left( \bm{m} \!\cdot\! \frac{\partial F_{0}}{\partial \bm{J}} \right)\, D_{\bm{m}} (\bm{J}) \, .
\label{definition_flux_total}
\end{equation}
In this expression, the sum on the resonances $\bm{m}$ will be restricted to the ILR, OLR and COR resonances. From the contours maps of $\bm{\mathcal{F}}_{\rm tot}$, we are able to explain the ridge observed in  S12 experiment.

One may rewrite the diffusion flux from equation~\eqref{definition_flux_total} using the \textit{slow} and \textit{fast} actions \citep{Lynden1979,Lynden1996}. We consider a given resonance $\bm{m}$ and introduce the change of coordinates
\begin{equation}
 J_{\bm{m}}^{s} = \frac{\bm{J} \!\cdot\! \bm{m}}{|\bm{m}|}  \, ,
\quad
 J_{\bm{m}}^{f} = \frac{\bm{J} \!\cdot\! \bm{m}^{\perp}}{|\bm{m}|}  \, ,
\label{definition_fast_actions}
\end{equation}
where $J_{\bm{m}}^{s}$ and $J_{\bm{m}}^{f}$ are respectively the slow and fast actions associated to the resonance ${\bm{m} = (m_{r},m_{\phi})}$, ${\bm{m}^{\perp} = (m_{\phi},- m_{r})}$ corresponds to the direction perpendicular to the resonance, and ${|\bm{m}| = \sqrt{\bm{m}\!\cdot\! \bm{m}}}$. Using the chain rule, for any function $X(J_{\phi} , J_{r})$, one has
\begin{equation}
\bm{m} \!\cdot\! \frac{\partial X}{\partial \bm{J}} = |\bm{m}| \, \frac{\partial X}{\partial J_{\bm{m}}^{s}} \bigg|_{J_{\bm{m}}^{f} = cst.} \!\!\!\!.
\label{chain_rule}
\end{equation}
We also naturally introduce the vector basis elements $({\bm{e}_{\bm{m}}^{s} = \bm{m}/|\bm{m}|} , {\bm{e}_{\bm{m}}^{f} = \bm{m}^{\perp} / |\bm{m}|})$ associated with this change of coordinates. In order to ease our qualitative description, we will assume that in the flux, equation~\eqref{definition_flux_total}, only one specific resonance $\bm{m}$ dominates. The diffusion flux $\bm{\mathcal{F}}_{\bm{m}}$, associated to the resonance $\bm{m}$, expressed within this rotated basis $(J_{\bm{m}}^{s} , J_{\bm{m}}^{f})$, can now be rewritten under the form
\begin{equation}
\bm{\mathcal{F}}_{\bm{m}} (J_{\bm{m}}^{s} , J_{\bm{m}}^{f}) = |\bm{m}|^{2} D_{\bm{m}} (\bm{J}) \frac{\partial F_{0}}{\partial J_{\bm{m}}^{s}} \, \bm{e}_{\bm{m}}^{s} \, .
\label{flux_fast_action}
\end{equation}
This rewriting shows that as soon as only one resonance dominates the secular dynamics, the diffusion flux will be aligned with this resonance, causing a narrow mono-dimensional diffusion in the preferential $J_{\bm{m}}^{s}-$direction. During this secular diffusion, particles conserve their fast action $J_{\bm{m}}^{f}$, which can henceforth be considered as adiabatically invariant, whereas their slow action $J_{\bm{m}}^{s}$ gets to change. We will show  that such a mono-dimensional diffusion is indeed responsible for the ridge observed in S12 simulation.

The evolution of a real disc is a combination of such resonant diffusions.
Because stars are born on circular orbits, action space is at first mostly populated close to the $J_{\phi}-$axis. Diffusion in the $J_{r}-$direction tends to increase the velocity dispersion within the disc and \textit{heats} it. Diffusion in the $J_{\phi}-$direction brings stars from one nearly circular orbit to another of different radius and is called radial migration. This mechanism does not heat the disc, and because the density of stars does not change rapidly along the $J_{\phi}-$axis, it can 
go unnoticed. However, chemical evolution within the disc induce a radial gradient in metallicity with which radial migration can interact \citep{SellwoodBinney2002}. Indeed, near the Sun, it tends to wipe out the correlation between the ages and metallicites of stars, by bringing to the Sun both old metal-rich stars formed at smaller radii and young metal-poor stars formed at larger radii.

%
\begin{figure}
\begin{center}
\epsfig{file=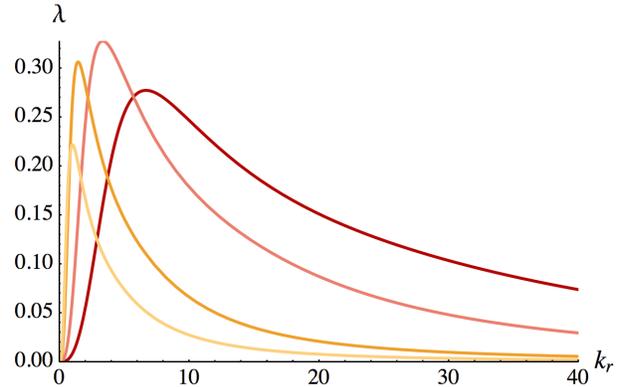,angle=-00,width=0.45\textwidth}
\caption{\small{Variations of the response matrix eigenvalues $\lambda$ with the WKB-frequency $k_{r}$. Red = small $J_{\phi}$, Yellow = Large $J_{\phi}$.}}
\label{figLambdakr}
\end{center}
\end{figure}

\subsection{The WKB tepid disc approximation}

One of the main difficulty of the implementation of the secular diffusion equation~\eqref{diffusion_equation} is to evaluate the diffusion coefficients, equation~\eqref{expression_diffusion_coefficients_all}. This difficulty remains even within the assumption of a simple coloured noise, as introduced in equation~\eqref{assumption_noise}. Indeed, it requires to define explicitly potential basis elements $\psi^{(p)}$, so that one can determine the response matrix from equation~\eqref{Fourier_M}. The next step is to invert this response matrix $\widehat{\mathbf{M}}$, in order to compute the diffusion coefficients $D_{\bm{m}} (\bm{J})$ from equation~\eqref{expression_diffusion_coefficients_all}. To ease these calculations, one may rely on the WKB assumption \citep{WKB,Toomre1964,Kalnajs1965,Lin1966}, which assumes that the perturbations will take the form of tightly wound spirals, which in turn allows us to write Poisson's equation \textit{locally}. Such an approximation is well-suited to study S12's experiment, because the secular evolution therein is sustained by the spontaneous generation of transient spirals. Considering only such perturbations amounts to considering basis elements with specific properties and shapes. 
 As explained in Appendix~\ref{sec:WKB_approximation}, the main consequence of the WKB approximation is that the response matrix from equation~\eqref{Fourier_M} becomes diagonal. The expression of its eigenvalues, $\lambda_{k_{\phi},k_{r},R_{0}} (\omega)$,
  for a tepid disc reads
\begin{equation}
\lambda_{k_{\phi},k_{r}R_{0}} (\omega) = \frac{2 \pi G \xi \Sigma_{\rm t} |k_{r}|}{\kappa^{2} (1 - s^{2})} \mathcal{F} (s,\chi) \,,
\label{expression_eigenvalues}
\end{equation}
where we have taken into account that only a fraction $\xi$ of the disc is self-gravitating. Here $k_{r}$ corresponds to the radial frequency of the local spiral perturbation which is getting amplified, $k_{\phi}$ is its azimuthal coefficient, which verifies $k_{\phi} = 2$ for the ILR, OLR and COR, and $R_{0}$ is the radius at which $\kappa$, $\Sigma_{t}$ and $\chi$ have to be evaluated. Here $s$ is a dimensionless parameter reading
\begin{equation}
s = \frac{\omega - k_{\phi} \, \Omega_{\phi}}{\kappa} \,.
\label{definition_s}
\end{equation}
The dimensionless quantity $\chi$ is given by
\begin{equation}
\chi = \frac{\sigma_{r}^{2} \, k_{r}^{2}}{\kappa^{2}} \,.
\label{definition_chi}
\end{equation}
Finally, the reduction factor, $\mathcal{F} (s, \chi)$, \citep{Kalnajs1965,Lin1966} is defined as
\begin{equation}
\mathcal{F} (s,\chi) = 2 \, (1 \!-\! s^{2}) \frac{e^{- \chi}}{\chi} \sum\limits_{m_r = 1}^{+ \infty} \frac{\mathcal{I}_{m_r} [\chi]}{ 1 - \big[ \frac{s}{m_r}\big]^{2}} \, .
\label{definition_fonction_F}
\end{equation}

Within the WKB approximation, one can show that the diffusion coefficients (fed by a stationary external perturbation ${\psi^{e} \propto \Sigma_{\rm t}^{1/2}}$ depending only on  position $J_{\phi}$, see equation~\eqref{assumption_noise}) can be expressed in terms of the response eigenvalues under the form
\begin{equation}
D_{\bm{m}} (\bm{J}) =  \Sigma_{\rm t} (R_{g}) \int \! \mathrm{d} k_{r} \, \mathcal{J}_{m_r}^{2} \!\bigg[\! \sqrt{\frac{2 J_{r}}{\kappa}} k_{r} \!\bigg]  \bigg[ \frac{1}{1 \!-\! \lambda_{k_{r}}} \bigg]^{2} \,,
\label{diffusion_coefficients_WKB}
\end{equation}
where $\mathcal{J}_{m_{r}}$ is the Bessel function of the first kind of index $m_{r}$ and the integration on $k_{r}$ corresponds to an integration on all the radial frequencies of the tightly wound spirals, each one being amplified by the amplification factor $1 / (1 \!-\! \lambda_{k_{r}})$. 
Equation~(\ref{diffusion_coefficients_WKB}) is a novel result derived in  Appendices A and C (see also paper I). 
The eigenvalues $\lambda_{k_{r}}$ have to be evaluated at the resonances so that $\omega = \bm{m} \!\cdot\! \bm{\Omega}$. At such a resonance, one can see that for $\bm{m} \!=\! (m_{r} , m_{\phi})$, $s$ is given by $s \!=\! m_{r}$. In order to handle the singularity of the  equation~\eqref{expression_eigenvalues} appearing for $s \!=\! \pm 1$, one adds a small imaginary part to the frequency of evaluation, so that $s \!=\! m_{r} + i \eta$. Indeed, as long as $\eta$ in modulus is small compared to the imaginary part of the least damped mode of the disc, adding this complex part has a negligible contribution on the expression of $\text{Re} (\lambda)$. Finally, we notice that for the Fourier coefficients associated to the inner/outer Lindblad resonances, one has $m_{r}^{\rm ILR} \!=\! -1$ and $m_{r}^{\rm OLR} \!=\! 1$. As  equation~\eqref{expression_eigenvalues} only depends on $s^{2}$, these two resonances have the same response matrix eigenvalues, and as we have assumed that they are subject to the same noise, will therefore lead to diffusion coefficients of equal magnitude.
\begin{figure}
\begin{center}
\epsfig{file=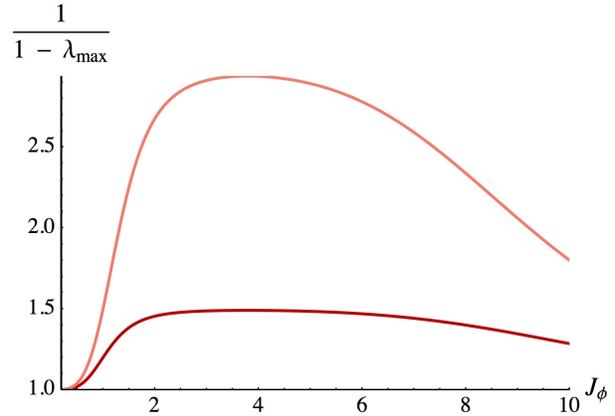,angle=-00,width=0.45\textwidth}
\caption{\small{Dependence of the amplification factor ${ {1}/{(1 \!-\! \lambda_{\rm max})} }$ with the position $J_{\phi}$ for the ILR resonance (red) and the COR resonance (pink). Throughout the entire disc, the COR resonance is more amplified than the ILR and OLR resonances}}
\label{figLambdaJphi}
\end{center}
\end{figure}
\subsection{Properties of the WKB equation}

Let us first study the behavior of the function ${k_{r} \mapsto \lambda_{k_{r}}}$ for a given resonance,
 $\bm{m}$, and angular momentum, $J_{\phi}$. This function describes at a given position how much a perturbation with a frequency $k_{r}$ is locally amplified. Figure~\ref{figLambdakr} shows that for a given angular momentum $J_{\phi}$, there is a preferred frequency $k_{r}^{\rm max} (J_{\phi}) $ for which $\lambda (k_{r})$ is maximum.
One can now simplify the expression of the diffusion coefficients from equation~\eqref{diffusion_coefficients_WKB} thanks to the behavior of the function $k_{r} \mapsto \lambda_{k_{r}}$. Indeed, one can see that this function is peaked around ${k_{r}^{\rm max} (J_{\phi})}$ with a typical spread equal to $\Delta k_{\lambda} (J_{\phi})$. Considering only the contribution from the region where $\lambda_{k_{r}}$ is maximum, equation~\eqref{diffusion_coefficients_WKB} becomes
\begin{equation}
\begin{aligned}
\hskip -0.3cm
D_{\bm{m}} (\bm{J}) \!=\! \, \Sigma_{\rm t} (R_{g}) \, \Delta k_{\lambda} \, \mathcal{J}_{m_{r}}^{2} \!\bigg[\! \sqrt{\frac{2 J_{r}}{\kappa}} \, k_{\rm max} \!\bigg] \bigg[\! \frac{1}{1 \!-\! \lambda_{\rm max}} \!\bigg]^{2} .
\end{aligned}
\label{diffusion_coefficients_small_denominators}
\end{equation}
The typical width of the amplification peak $\Delta k_{\lambda}$ is estimated via the width at half-maximum of the function ${k_{r} \mapsto \lambda_{k_{r}}}$, so that ${\Delta k_{\lambda} = k_{1/2}^{\rm sup} \!-\! k_{1/2}^{\rm inf}}$, where $k_{1/2}^{\rm sup}$ and $k_{1/2}^{\rm inf}$ are the two solutions of the equation ${\lambda (k_{r}) = \lambda_{\rm max} / 2}$. For the specific case of a Mestel disc considered by S12, the spread $\Delta k_{\lambda}$ and its position $k_{\rm max}$ satisfy an additional property. Indeed, we have supposed that throughout the disc the radial-action spread $\sigma_{r}^{2}$ was constant and we know from equation~\eqref{expression_intrinsic_frequencies} that the epicyclic frequency $\kappa$ varies as $1 / J_{\phi}$. As a consequence, one has from equation~\eqref{definition_chi} that ${\chi \propto (k_{r} J_{\phi})^{2}}$. One may then rewrite the dependence with $k_{r}$ and $J_{\phi}$ of the eigenvalues $\lambda$ under the form: ${\lambda (k_{r} , J_{\phi}) = f_{1} (J_{\phi}) \, f_{2} (k_{r}  J_{\phi})}$, where $f_{1}$ and $f_{2}$ are given functions, which depend on the resonance considered. Such a dependence of $\lambda$ with the radial frequency $k_{r}$ immediately implies that an additional scale-invariance property is satisfied so that $\Delta k_{\lambda} (J_{\phi}) \propto 1/ J_{\phi}$ and $k_{\rm max} \propto 1/ J_{\phi}$. The inner regions of the disc have therefore larger eigenvalues spread than the outer regions. Such a dependence tends to enhance the susceptility of the most inner regions, which physically makes sense.

Another important factor in the diffusion coefficientsgiven by equation~\eqref{diffusion_coefficients_small_denominators} is the local amplification factor $1/(1 \!-\! \lambda_{\rm max})^{2}$, which describes the strength of the amplification due to the  self-gravity of the system. As noted previously, the ILR and OLR resonances possess the same amplification factor. However, one can compare the strength of the amplification for the ILR and the COR resonance as seen on figure~\ref{figLambdaJphi}.
One can note that the maximum amplification $1/(1 \!-\! \lambda_{\rm max})$ ($\sim 3$ for the COR and $\sim 1.5$ for the ILR) remain rather small so that the susceptibility of the disc is not too important. Moreover, one can note that the corotation resonance is more amplified than the inner/outer Lindblad resonances. 
However, what really represents the strength of the diffusion in the action diffusion map of the distribution is not the value of the diffusion coefficients $D_{\bm{m}} (\bm{J})$ but the flux given by $D_{\bm{m}} (\bm{J}) \, \bm{m} \!\cdot\! ( \partial F_{0} / \partial \bm{J} )$. As the disc is tepid, one has for most of the regions $\left| \partial F_{0} / \partial J_{r} \right| \gg  \left| \partial F_{0} / \partial J_{\phi} \right|$, so that the ILR and OLR resonances for which $m_{r} \neq 0$ are favoured by the inhomegeneity of the distribution function compared to the COR resonances. This arbitration between the inhomogeneity of the disc and its susceptibility determines the dominant regime of secular diffusion undergone by the disc.

\begin{figure*}
\begin{center}
\epsfig{file=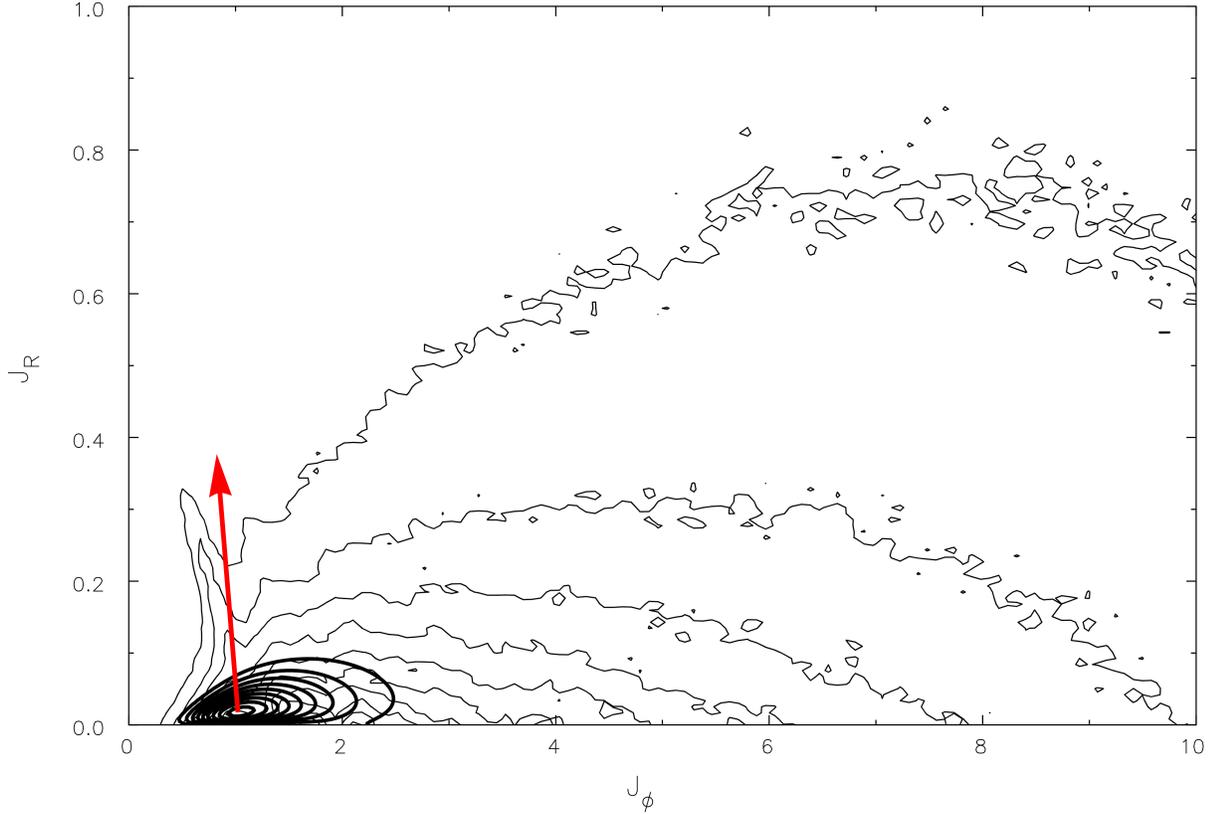,angle=-00,width=0.90\textwidth}
\caption{\small{Map of the norm of the total flux summed for the three resonances (ILR, COR, OLR) (\textit{bold lines}). The contours are spaced linearly between 95\% and 5\% of the function maximum. The red vector gives the direction of the vector flux associated to the norm maximum (arbitrary length). The background  contours correspond to the \textit{diffused} distribution from S12 (\textit{thin lines}), which exhibits a narrow ridge of diffusion.}}
\label{figNorm_Flux}
\end{center}
\end{figure*}
%

\section{Results}
\label{sec:results}

In order to explain the ridge observed in S12, one can now compute on the plane $(J_{r} , J_{\phi})$ the value of the diffusion coefficients $D_{\bm{m}}(\bm{J})$ from equation~\eqref{diffusion_coefficients_small_denominators} for the three main resonances and obtain a numerical estimation of the maximum flux from equation~\eqref{definition_flux_total}, from which the secular diffusion will start. 
We will then vary the main features of the galaxy and its environment to trace its effects on the ridge.

\subsection{Reproducing the S12 experiment}
\label{sec:S12results}

The results using the numerical prefactors from S12 are presented on figure~\ref{figNorm_Flux}.
The thin lines  represents the contours of the \textit{diffused} distribution function obtained by S12, which exhibits a narrow ridge of diffusion. Superimposed on these contours are shown the contours of the norm of the secular diffusion flux from equation~\eqref{definition_flux_total}, and the direction associated to this flux.

There is only one maximum peak of diffusion located in ${(J_{r} , J_{\phi}) \simeq (0.01\,,\,1)}$ from which the secular diffusion will unambiguously start. One should note that the predicted position of the peak of diffusion is slightly offset from the one observed in S12, which was around ${J_{\phi} \simeq 1.2}$. This difference may be due to our  crude model of intrinsic noise from equation~\eqref{assumption_noise}, 
the fact that we are comparing the initial diffusion direction to the \textit{final} position of the ridge (see section~\ref{ref:varying-temperature} below), 
and/or possibly also to the limitations of the WKB approach which is only approximately accurate in a regime where the transients spirals are not sufficiently tightly wound. However, even so, the agreement on the regime of secular diffusion undergone by the disc remains quite good. To this maximum of the norm of the diffusion flux is also associated a direction of diffusion. The direction of the ILR-diffusion is associated to the vector $(-2,1)$ in the $(J_{\phi} , J_{r})-$plane, which makes an angle of $153^{\circ}$ with the $J_{\phi}-$axis. The direction of diffusion predicted within our approach is of approximately $120^{\circ}$. This quasi-alignement illustrates the fact that the ILR is the main resonance of the secular evolution of the tapered Mestel disc. S12 had found that the diffusion in action space was completely dominated by the ILR resonance, so that the ridge was aligned with the direction $\bm{m} = \bm{m}_{\rm ILR}$. In our case, the slight misalignement observed has two origins. First of all, we considered a total secular flux, equation~\eqref{definition_flux_total}, summed on the three resonances ILR, OLR and COR, which all have different diffusion directions, so that it tends to slightly rotate the direction of the dominant resonance.

 Moreover, recall that our noise assumption from equation~\eqref{assumption_noise}  has no ${ \omega = \bm{m} \!\cdot\! \bm{\Omega} }$ dependence, so that the ILR and OLR resonances possess the same susceptibility, which leads to an over-representation the OLR resonance. However, we unambiguously recover that the S12 disc was in a ILR dominated regime, taking place in the inner regions of the disc. Since the ILR dominates the diffusion, let us use the analysis of mono-dimensional diffusion. The slow and fast actions associated to the ILR resonance are given by ${J_{\rm ILR}^{s} \propto J_{\phi} \!-\! J_{r} / 2}$ and ${J_{\rm ILR}^{f} \propto J_{\phi} / 2 \!+\! J_{r}}$. In the neighbourhood of the diffusion peak, stars can therefore be assumed to diffuse along lines of constant $J_{\rm ILR}^{f}$, along which their slow action $J_{\rm ILR}^{s}$ changes. This leads to an mono-dimensional diffusion causing an \textit{heating} of the disc.
 
  Also,  note from  figure~\ref{figNorm_Flux} that the diffusion flux norm is non-negligible only in a narrow band in $J_{\rm ILR}^{s}$. The size of this region will determine the typical width of the narrow ridge in the $J_{\rm ILR}^{f}-$direction observed in S12 simulations. Starting from a narrow region in $J_{\rm ILR}^{f}$ and diffusing predominantely in the $J_{\rm ILR}^{s}-$direction, one can therefore explain the ridge of limited extent in the $(J_{\phi},J_{r})-$plane observed numerically in S12, and correctly captured by the WKB limit of the secular diffusion formalism.
  
   This secular behavior of a typical Mestel disc dominated by the ILR resonance in the inner regions of the disc can be interpreted in the following way. Recall that the intensity of the secular diffusion is encoded by the total flux from equation~\eqref{definition_flux_total}.  The use of a fraction, $\xi = 0.5$, for the surface density reduced the susceptibility of the disc and therefore reduced the amount by which perturbations can be amplified through self-gravity. Consequently, the susceptibility-structure of the disc via $D_{\bm{m}} (\bm{J})$ is not the only key parameter to determine the peaks of diffusion, but rather its inhomogeneous structure represented by the gradients of the distribution function ${ \partial F_{0} / \partial \bm{J} }$. This has two implications. Since the disc is tepid, the distribution function gradients are more important for the ILR and OLR resonances than for the COR, as resonances with a non-zero $m_{r}$ component are favored. Moreover, the gradients are the highest for the inner-regions, where the cut-off takes place, because of the presence of the tapering function ${ T_{\rm inner} (J_{\phi}) }$. This clarifies why it could have been expected that the secular diffusion would predominantly take place through a $J_{r}-$heating in the inner regions.

\begin{figure*}
\begin{center}
\begin{tabular}{@{}ccc@{}}
\epsfig{file=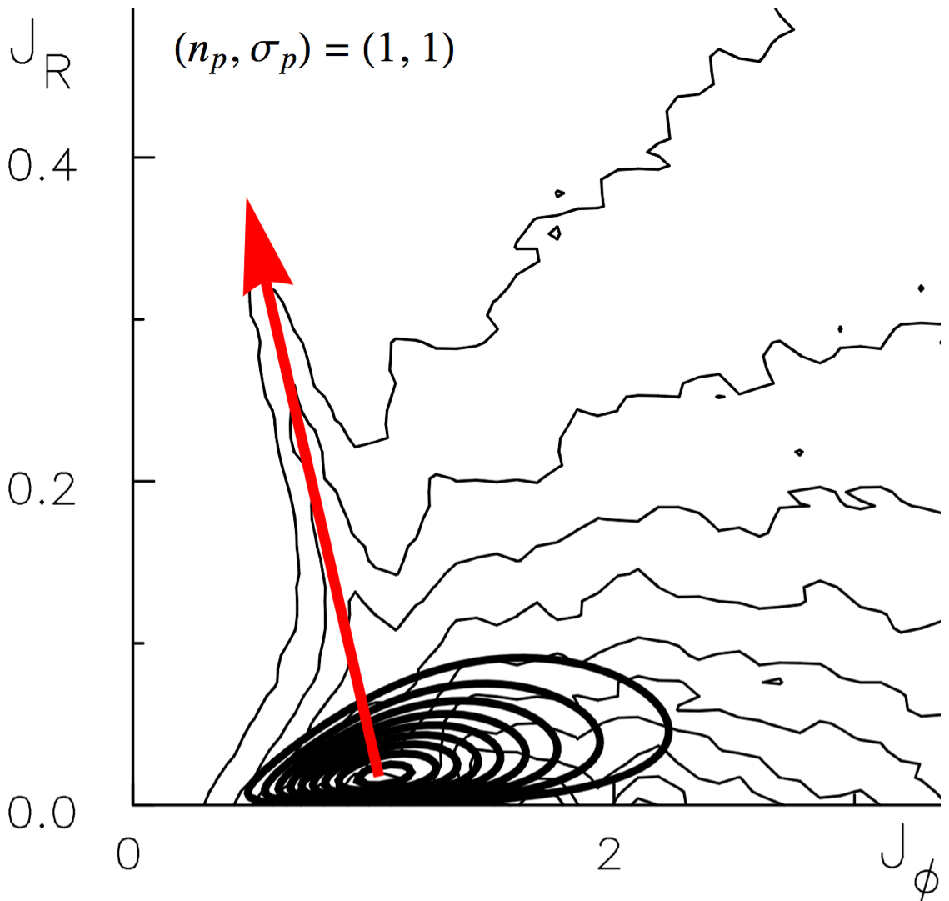,angle=-00,width=0.30\textwidth} &
\epsfig{file=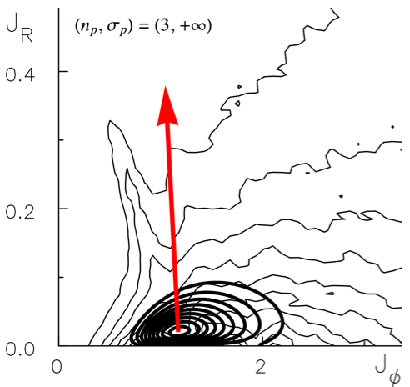,angle=-00,width=0.30\textwidth} &
\epsfig{file=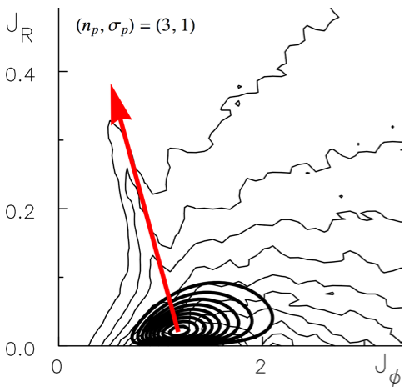,angle=-00,width=0.30\textwidth}
\end{tabular}
\caption{\small{Dependence of the maximum peak of diffusion and its associated direction with the (ad hoc) properties of the perturbations of the system given by equation~\eqref{assumption_noise_adhoc}. 
The various curves follow figure~\ref{figNorm_Flux}.
From left to right: $(n_{p},\sigma_{p}) = (1\,,\,1) \,,\; (3\,,\, + \infty) \,,\; (3\,,\,1)$. Increasing the power index $n_{p}$ tends to shift the peak position to higher $J_{\phi}$, whereas decreasing $\sigma_{p}$ enhances the importance of the ILR and tends to align the direction of diffusion with $\bm{m}_{\rm ILR}$.}}
\label{figNorm_Flux_adhoc}
\end{center}
\end{figure*}

An additional feature of such ILR-dominated diffusion is the typical temporal growth rate of the ridge in action space. As shown in Appendix~\ref{sec:ridge_growthrate} such an anisotropic and mono-dimensional diffusion, when started from a narrow hot spot, leads to a \textit{faster} diffusion than the usual homogeneous heat equation. Indeed, the secular heating of a galactic disc through the ILR-resonance corresponds to a \textit{super-diffusive} case, for which the diffusion of the distribution function diffuses does not follow the usual growth rate in $\sqrt{t}$ of the homogeneous heat equation: equation~\eqref{temporal_growth_rate_ILR} states that the scattering of the ridge in the $J_{\rm ILR}^{s}-$direction grows linearly with the time $t$.

\subsection{Modifying the properties of the perturbation}
\label{subsec:perturb}

The spectral properties of the noise play an important role in the detailled properties of the diffusion. In order to qualitatively describe this impact, one can slightly modify the assumption~\eqref{assumption_noise} on the noise and consider the more general class of perturbation
\begin{equation}
\left< \left| \psi^{\rm ext} (J_{\phi}, \omega) \right|^{2}\right> \simeq \bigg[ \Sigma_{\rm t} (J_{\phi}) \bigg]^{n_{p}} \!\!\exp\! \bigg[\! - \frac{\omega}{\sigma_{p} \, \Omega (J_{\phi})} \!\bigg] \,,
\label{assumption_noise_adhoc}
\end{equation}
where $n_{p}$ is a power index introduced in order to modify the assumption on the Poisson shot noise, and $\sigma_{p}$ is a dimensionless parameter (rescaled thanks to $\Omega (J_{\phi})$) adding a dependence on the temporal frequencies $\omega$.  Our initial assumption, equation~\eqref{assumption_noise} corresponds to the case ${n_{p}=1}$ and ${\sigma_{p} \to + \infty}$. In order to illustrate the impact of the perturbation's power spectrum on the properties of the secular diffusion, possible ad hoc choices of the parameters ${(n_{p}, \sigma_{p})}$ are shown on figure~\ref{figNorm_Flux_adhoc}.
Increasing the index $n_{p}$ enhances the relative importance of the surface density, and from the figure~\ref{figSigmaDF}, one can see that it will favor the region around ${J_{\phi} \simeq 1.5}$, where the inner tapering takes place. The choices of the inner tapering function $T_{\rm inner}$ from equation~\eqref{inner_outer_taper_Sellwood} and its cutting scale $R_{i}$ are then responsible for the location of the maximum surface density on figure~\ref{figSigmaDF} and therefore for the position of the peak of maximum diffusion.
Moreover, adding a dependence on $\omega$ breaks the degeneracy between the ILR and OLR resonances. Indeed, the frequency associated to a resonance $\bm{m}$ is given by ${\omega = \bm{m} \!\cdot\! \bm{\Omega}}$. From  expression~\eqref{expression_intrinsic_frequencies} for the intrinsic frequencies, one can note that the frequencies of the three resonances satisfy the inequalities
\begin{equation}
0 < \omega_{\rm ILR} < \omega_{\rm COR} < \omega_{\rm OLR} \, .
\label{inequality_resonances_frequencies}
\end{equation}
 Consequently, the addition of the decaying exponential in the noise model~\eqref{assumption_noise_adhoc} tends to enhance the ILR resonance compared to the other resonances. In the sum in equation~\eqref{definition_flux_total} of the total flux, the vector contribution from the ILR dominates the other resonances, and we recover the fact that the direction of diffusion associated to the peak of secular diffusion is closely aligned with $\bm{m}_{\rm ILR}$ compared to what was observed on figure~\ref{figNorm_Flux}, with the simpler noise assumption given by equation~\eqref{assumption_noise}, for which the COR resonance plays a more important role. Such ad-hoc experiments illustrate how the detailled properties and dependence of the perturbations with $J_{\phi}$ and $\omega$ can impact the characteristics of the secular diffusion. 
 
One could imagine situations where matching the observed secular response of families of disc to a given model for the external 
perturbation would provide means of characterising the statistical properties of their cosmic environment.

\subsection{Modifying the tapering of the disc}

\begin{figure*}
\begin{center}
\begin{tabular}{@{}ccc@{}}
\epsfig{file=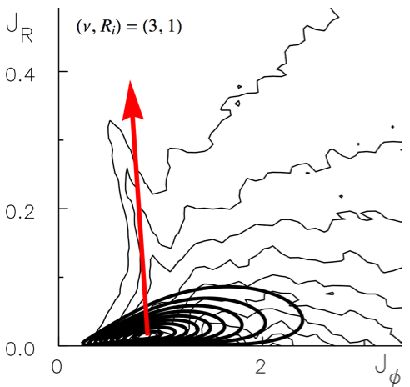,angle=-00,width=0.30\textwidth} &
\epsfig{file=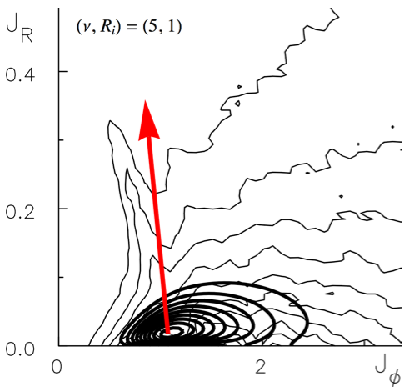,angle=-00,width=0.30\textwidth} &
\epsfig{file=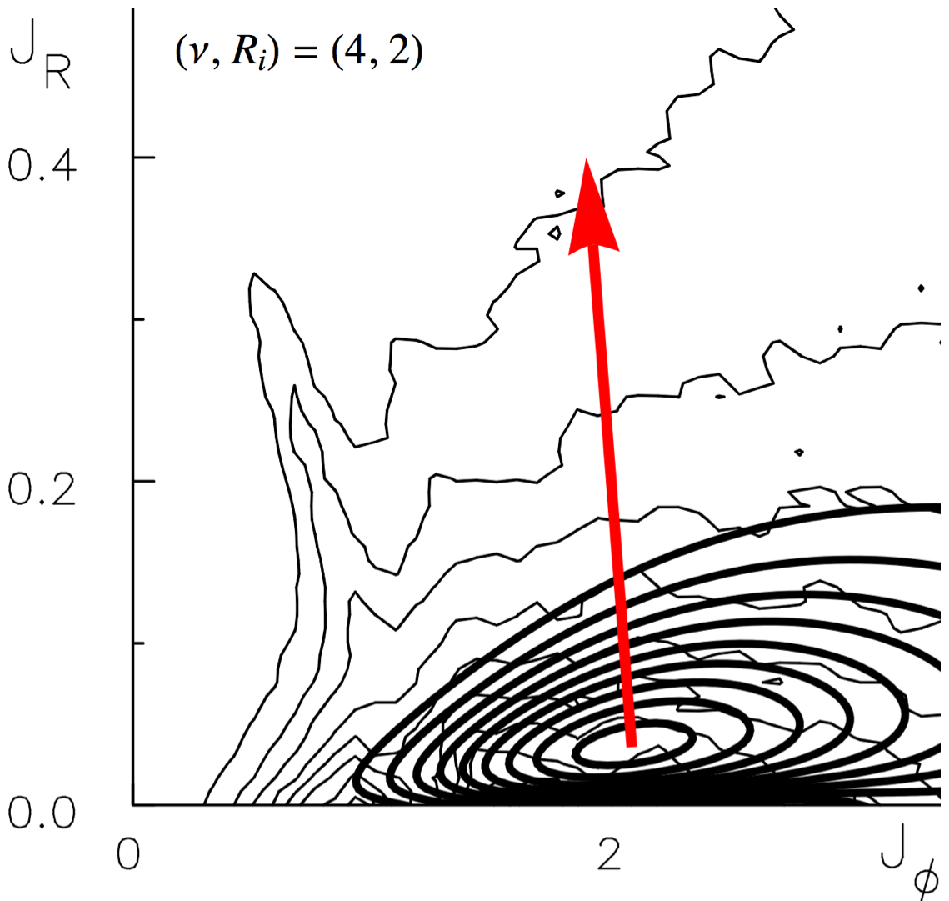,angle=-00,width=0.30\textwidth} 
\end{tabular}
\caption{\small{Dependence of the maximum peak of diffusion and its associated direction with the inner tapering function $T_{\rm inner}$ from equation~\eqref{inner_outer_taper_Sellwood}. 
The various curves follow figure~\ref{figNorm_Flux}. Here $T_{\rm inner}$ is characterized by the pair $(\nu, R_{i})$. The S12 case corresponds to ${(\nu , R_{i}) = (4\,,\,1)}$.
From left to right: ${(\nu , R_{i}) = (3\,,\,1) \,,\; (5\,,\,1) \,,\; (4\,,\,2)}$. Reducing $\nu$ reduces the sharpness of the tapering and therefore reduces the gradients of the distribution function so that the peak of diffusion migrates to the most inner regions. The scale radius $R_{i}$ is as expected a crucial parameter to determine the position of the peak maximum.}}
\label{figNorm_Flux_taper}
\end{center}
\end{figure*}

The inner tapering function $T_{\rm inner}$ from equation~\eqref{inner_outer_taper_Sellwood}, which represents the bulge of the galaxy, plays a crucial role to secularly induce an ILR dominated peak of diffusion in the inner regions. Here $T_{\rm inner}$ is characterized by two parameters $(\nu,R_{i})$, where $\nu$ controls the sharpness of the tapering, whereas $R_{i}$ is the scale at which it takes place. In order to illustrate the impact of $T_{\rm inner}$ on the secular diffusion, some modifications of $T_{\rm inner}$ are shown in figure~\ref{figNorm_Flux_taper}.
When $\nu$ is increased, the surface density $\Sigma_{\rm t}$ becomes steeper, so that the inhomogeneity of the system, represented by ${\partial F_{0} / \partial \bm{J}}$, becomes more important. As a consequence, the peak of diffusion tends to migrate to the region of higher DF gradients, which are in the vicinity of the scale radius $R_{i}$. On the other hand, decreasing $\nu$ tends to enhance the importance of the susceptibility coefficients $D_{\bm{m}} (\bm{J})$ in the determination of the peak of diffusions. We noted in equation~\eqref{diffusion_coefficients_small_denominators}, that the scale-invariance properties of the Mestel disc impose that ${\Delta k_{\lambda} \propto 1/J_{\phi}}$, so that the inner regions are naturally more susceptible. As a consequence, decreasing $\nu$ tends to migrate the peak of diffusion inwards. Finally, as expected, modifying the scale radius $R_{i}$ naturally leads to a similar displacement of the peak of diffusion. One should  note that such modifications do not have any significant impact on the direction of  diffusion.

\subsection{Modifying the temperature of the disc}
\label{ref:varying-temperature}
\begin{figure*}
\begin{center}
\begin{tabular}{@{}ccc@{}}
\epsfig{file=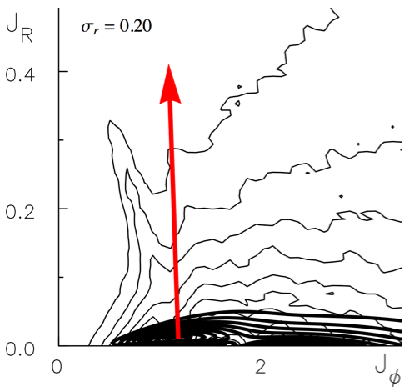,angle=-00,width=0.30\textwidth} &
\epsfig{file=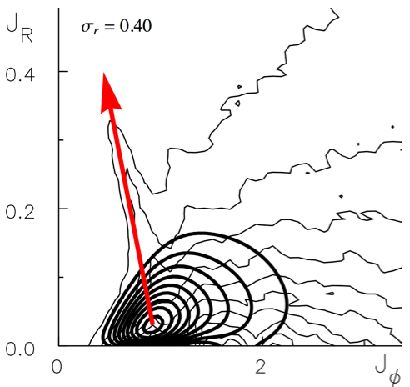,angle=-00,width=0.30\textwidth} &
\epsfig{file=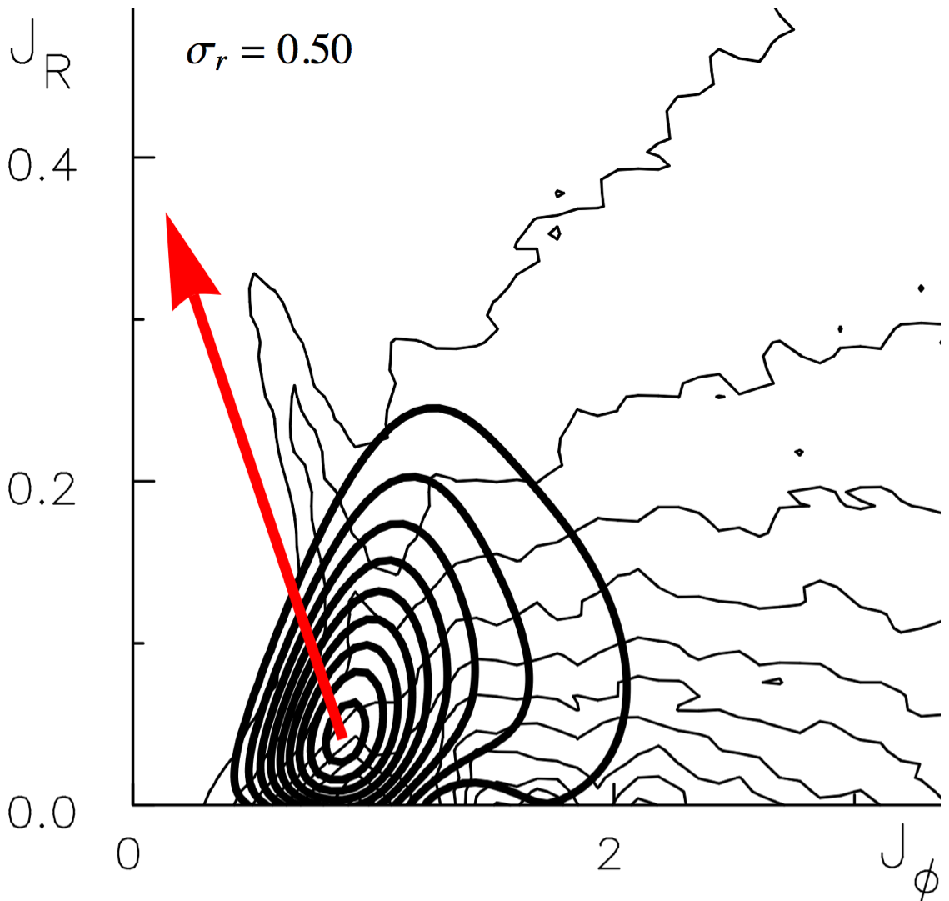,angle=-00,width=0.30\textwidth} 
\end{tabular}
\caption{\small{Dependence of the maximum peak of diffusion and its associated direction on the \textit{temperature} of the disc, encoded by $\sigma_{r}$. 
The various curves follow figure~\ref{figNorm_Flux}. The S12 case corresponds to ${ \sigma_{r} = 0.284 }$.
From left to right: $\sigma_{r} = 0.20 \,,\; 0.40 \,,\; 0.50$\,. Larger $\sigma_{r}$ corresponds to hotter disc and therefore more stable. Colder discs have a diffusion peak with a wider $J_{\phi}-$extension and therefore will lead to wider ridges.}}
\label{figNorm_Flux_sigmar}
\end{center}
\end{figure*}

An additional way to modify the property of the disc is to change its temperature by varying the value of $\sigma_{r}$, which encodes the radial-action spread of the distribution, as in equation~\eqref{DF_Jr_Jphi_isothermal}. Its impact is illustrated in figure~\ref{figNorm_Flux_sigmar}.
Decreasing $\sigma_{r}$ leads to colder discs, which tend to possess a wider peak of diffusion in regions slightly more external. A wider peak of diffusion will lead to a wider ridge when the secular diffusion will take place. The impact of $\sigma_{r}$ on the secular diffusion properties is in fact convolved, as it influences both the stability of the disc via $Q$, but also the detailed properties of amplification of the disc, through the expression of the eigenvalues in equation~\eqref{expression_eigenvalues} via $\chi$. Both the characteristics of the peak of diffusion and its direction of diffusion are therefore influenced by $\sigma_{r}$. 

Interestingly, as the ridge effectively increases locally the temperature of the disc (see Appendix~\ref{sec:ridge_growthrate}), Figure~\ref{figNorm_Flux_sigmar} suggests that it
will in turn rotate the ridge in the ILR direction found by S12. Hence we can  expect the discrepancy found on figure~\ref{figNorm_Flux_xi_all} to decrease 
in time through this process.

\subsection{Increasing the active fraction of the disc}
\label{subsec:activefrac}

\begin{figure}
\begin{center}
\epsfig{file=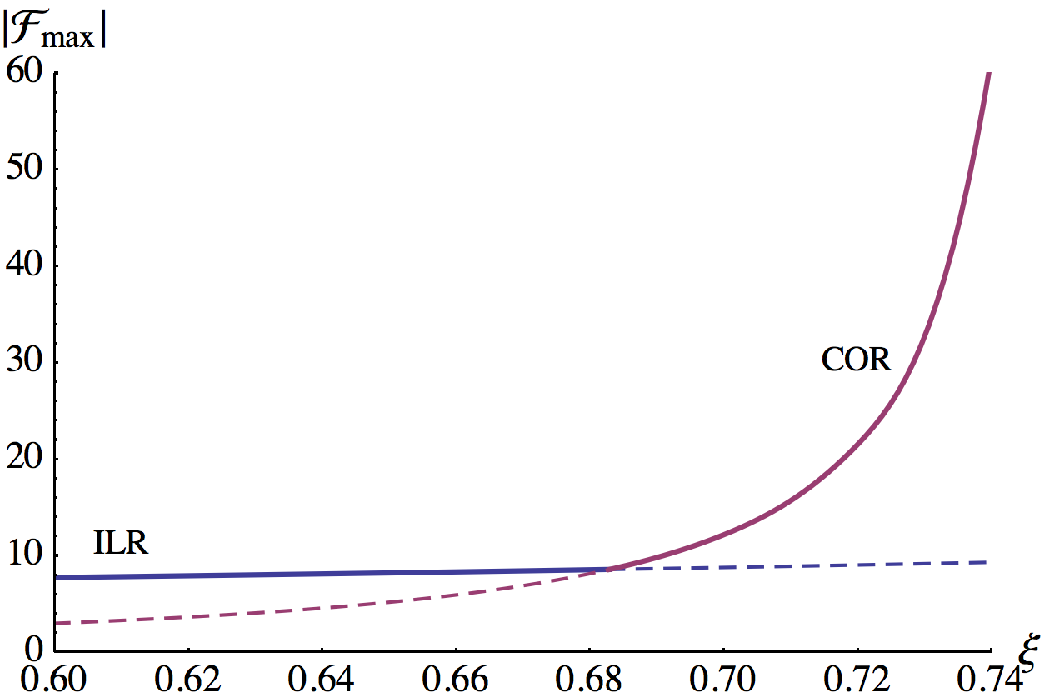,angle=-00,width=0.45\textwidth}
\caption{\small{For a given value of the active fraction $\xi$ of the disc, one can identify a peak of flux associated to the ILR resonance, around ${ (J_{r} , J_{\phi}) \simeq (0.01\,,\,1) }$,  and one to the COR resonance, close to ${ (J_{r} , J_{\phi}) \simeq (0\,,\,2) }$. This figure represents the dependence of the norm of the peak flux from equation~\eqref{definition_flux_total} for the two different regions of resonance, ILR (in \textit{blue}) and COR (in \textit{purple}), as the self-gravity of the disc is increased. For ${ \xi \geq 0.68 }$, the secular evolution of the disc becomes dominated by radial migration effects through the corotation resonance in the intermediate regions, rather than by an heating via the ILR resonance in the inner regions.}}
\label{figNorm_Transfert}
\end{center}
\end{figure}

\begin{figure*}
\begin{center}
\begin{tabular}{@{}ccc@{}}
\epsfig{file=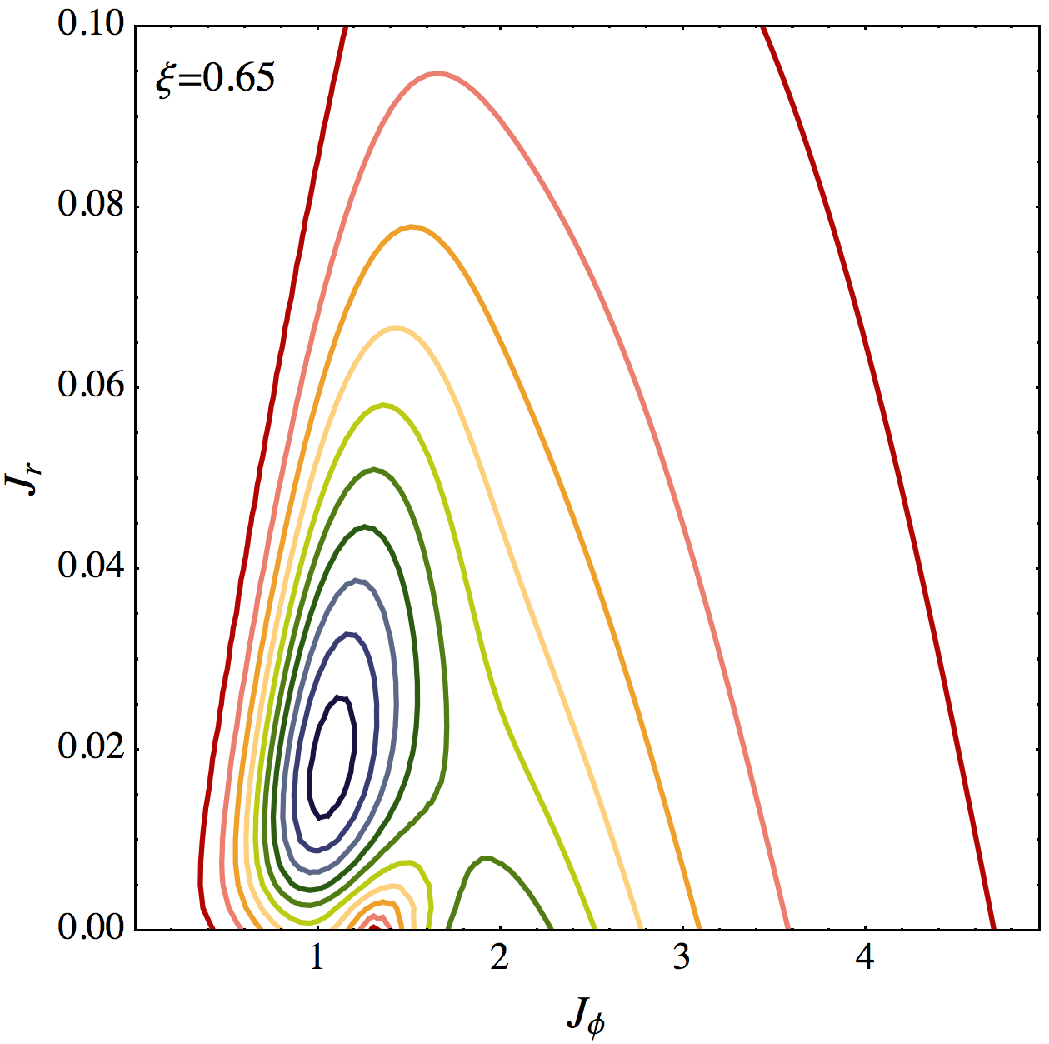,angle=-00,width=0.30\textwidth} &
\epsfig{file=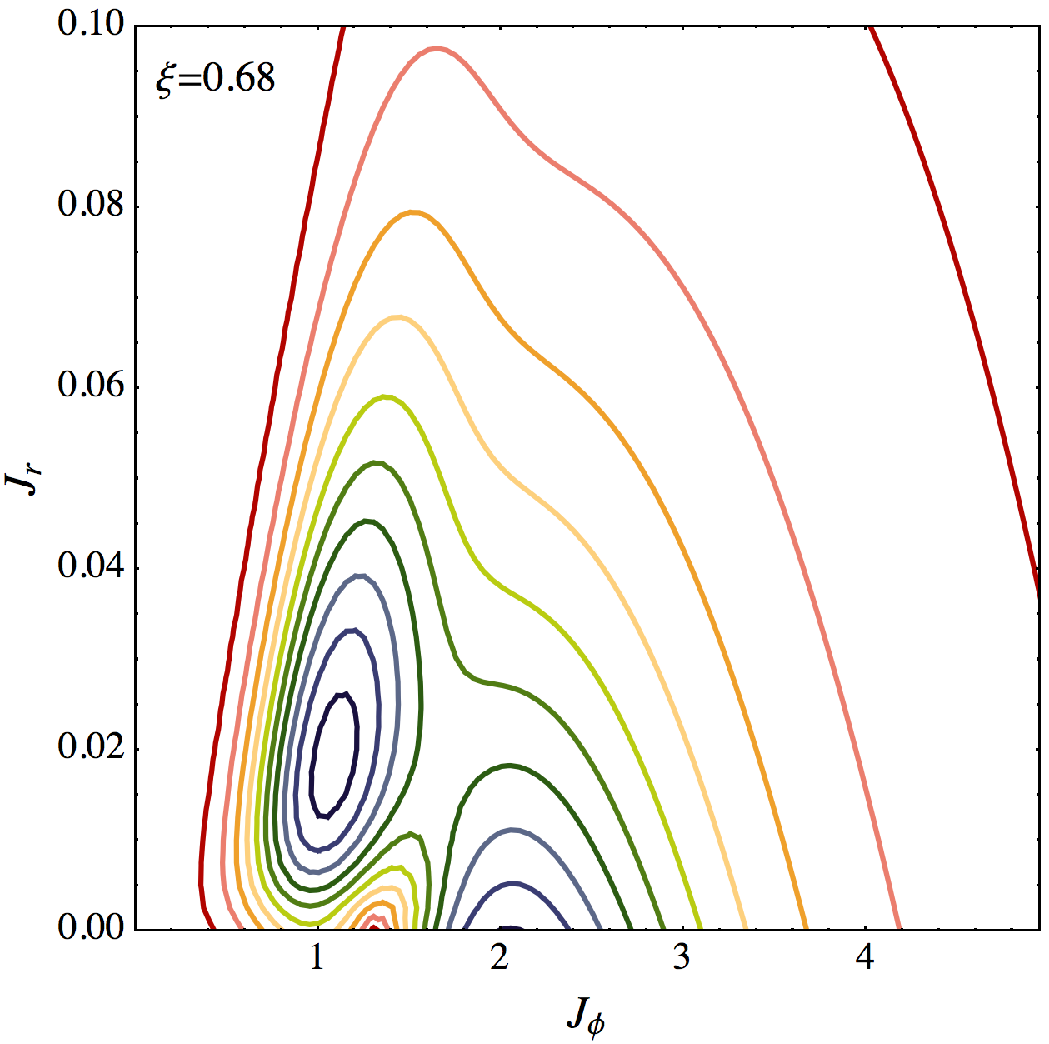,angle=-00,width=0.30\textwidth} &
\epsfig{file=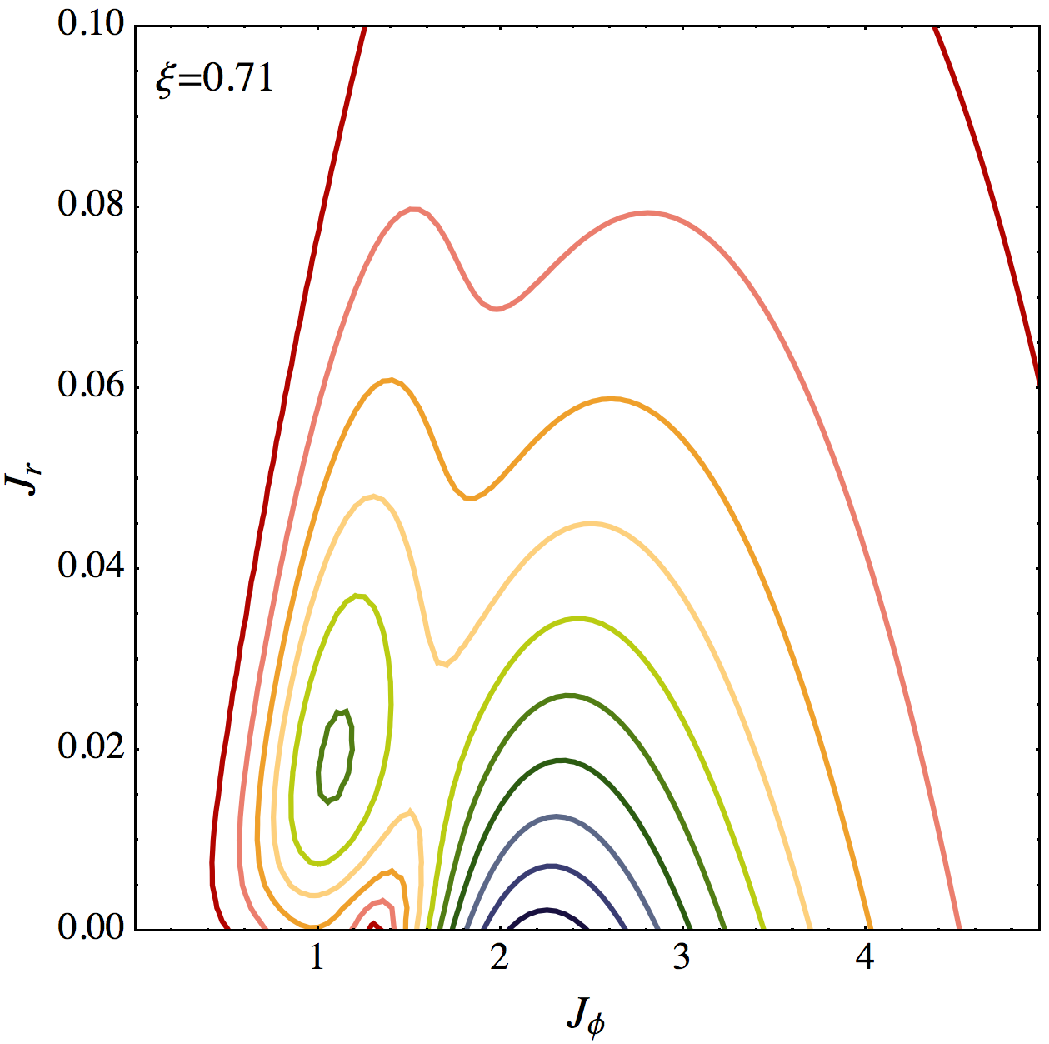,angle=-00,width=0.30\textwidth}
\end{tabular}
\caption{\small{Map of the norm of the  total flux summed for the three resonances (ILR, COR, OLR), when increasing the active fraction $\xi$ of the disc. The contours are spaced linearly between 95\% and 5\% of the function maximum for each case. From left to right: ${ \xi = 0.65 \,,\; 0.68 \,,\; 0.71 }$\,. (Such values of $\xi$ still comply with the constraint ${ Q(J_{\phi}) > 1 }$).}}
\label{figNorm_Flux_xi_all}
\end{center}
\end{figure*}

Let us now study  one last feature of diffusion while modifying some of the characteristics of the disc and the halo. The behaviour of the function ${J_{\phi} \mapsto 1/(1 \!-\! \lambda_{\rm max})}$ on figure~\ref{figLambdaJphi} showed that the COR resonance has higher amplification factors than the ILR and OLR resonances. In order not to be  dominated by the ILR resonance in the inner regions, one may try to increase the susceptibility of the disc so that the predominant diffusion will take place through the COR resonance. The eigenvalues of the response matrix from equation~\eqref{expression_eigenvalues} can be increased via  an increase of $\xi$, the active fraction of the disc. Figure~\ref{figNorm_Flux_xi_all} illustrates such changes.
One can note in figure~\ref{figNorm_Flux_xi_all} that as $\xi$ is increased, a significant COR-dominated region around the position ${(J_{r} , J_{\phi}) \simeq (0 \,,\, 2)}$ appears. This new diffusion region ends up being more important than the ILR-peak around the position ${(J_{r} , J_{\phi}) = (0.01 \,,\, 1)}$, as illustrated on figure~\ref{figNorm_Transfert}.
Indeed, as $\xi$ increases, both $\lambda_{\rm max}^{\rm ILR}$ and $\lambda_{\rm max}^{\rm COR}$ increases, but since one has ${\lambda_{\rm max}^{\rm ILR} \!<\! \lambda_{\rm max}^{\rm COR} \!<\! 1}$, for $\lambda_{\rm max}^{\rm COR}$ close to $1$, the COR resonance gets more amplified than the inner/outer Lindblad resonances. The fast and slow actions associated to the corotation resonance, are straightforwardly given by ${J_{\rm COR}^{s} \propto J_{\phi}}$ and ${J_{\rm COR}^{f} \!\propto\! J_{r}}$. As the fast action tends to be conserved during a secular diffusion dominated by only one resonance, we can conclude that the new peak of diffusion observed on figure~\ref{figNorm_Flux_xi_all} corresponds to the diffusion of nearly-circular orbits, which increase their angular momentum $J_{\phi}$, while their radial energy $J_{r}$ remains small. As the active fraction $\xi$ is  increased, we observe a transition from the ILR dominated heating of the inner region ($J_{\phi} \simeq 1$) to a regime of radial migration of quasi-circular orbits in more intermediate regions ($J_{\phi} \simeq 2$) as shown in figure~\ref{figNorm_Transfert}. With such higher active fractions, the secular diffusion mechanism is now in a different regime of long-term evolution, mainly determined by the susceptibility of the disc via the diffusion coefficients $D_{\bm{m}} (\bm{J})$, rather than by the inhomogeneity of the distribution function through $\partial F_{0} / \partial \bm{J}$.

\section{Conclusions}
\label{sec:conclusions}

The secular diffusion equation  \citep{Binney1988,weinberg93,Pichon2006} of a self-gravitating collisionless system was re-derived and implemented in the WKB limit, using angle-action variables for tightly wound spirals in a tepid disc described by a Schwarschild distribution function. In this limit, the functional form of the diffusion coefficient allowed us to identify the ridge found in action space by S12  for
a  stable Mestel disc. It  originates from a resonant mono-dimensional diffusion, which is maximum at a specific locus in the inner regions of the disc.
As the disc model investigated by S12 was somewhat singular,  its global scale invariance  is only broken by the addition of the inner and outer tapering functions (representing the bulge and the edge of the disc), which, as expected, also play an important role in the determination of the regime of secular evolution, hence the position of the peak of diffusion. 
 The birth of a resonant ridge is therefore the result of a subtle \textit{fine tuning} between many parameters of the system. Indeed, having for example a noise of the form ${\psi^{\rm ext} \!\propto \delta_{\rm D} (\omega \!-\! \omega_{p})}$, corresponding to a perturbation peaked at a specific frequency is not a mandatory condition to observe a resonant ridge. The self-gravity of the disc (via $\xi$ and $\lambda$), its susceptibility (via $D_{\bm{m}} (\bm{J})$), its inhomogeneity (via ${\partial F_{0} / \partial \bm{J}}$), its temperature (via $\sigma_{r}^{2}$), its bulge structure (via $T_{\rm inner}$), and the source of perturbations (via $\psi^{\rm ext}$), all play a non negligible role in the creation and the properties of the resonant ridge, as shown in the ad-hoc experiments of the sections \ref{subsec:perturb} to \ref{subsec:activefrac}.

The solar neighbourhood shows at least three indications of the secular mechanism described in this paper:
i) Groups of stars of a given age see their random velocities increase with the age of the group \citep{Wielen1977,AumerBinney2009}.
ii) Around the Sun, the velocity distribution of stars is made of various \textit{streams} of stars \citep{Dehnen1998}. Despite the fact that each stream is made of stars with different ages and chemistries, they respond to some perturbation in a similar way.
iii) In the $(J_{\phi},J_{r})$ space, one observes a depression of the density of stars near ${J_{r} = 0}$ and an enhancement for larger $J_{r}$, so that the disturbance in stellar density follows a curve consistent with resonant conditions \citep{McMillan2011}.
Given a detailed characterisation of the perturbations induced by e.g. the cosmic environment, one could study their effects on a typical self-gravitating collisionless galactic disc. In the context of the ongoing \textsc{GAIA} mission, this externally induced secular evolution is thought to be potentially a powerful approach to describe the long-term resonant radial migration of stars and its impact on the observed metallicity gradients \citep{SellwoodBinney2002,Roskar2008,Schonrich20092,Solway2012,Minchev2013}. 
 
More generally, the formalism of \textit{dressed} secular diffusion \textit{fed} by an exterior perturbations can be applied to any integrable model, and may lead to various secular diffusion scenarii depending on the structure of the galaxy and the properties of the spectrum of the external perturbations. 
The WKB formulation, when applicable, is very useful to yield a multiplicative amplification of the exterior perturbation and a tractable \textit{quadrature} for the diffusion coefficients, which under certain circumstances can be written algebraically (equation~\eqref{diffusion_coefficients_small_denominators}).  
Such simplification provides useful  insight into the  physical processes at work, e.g. the relevant resonances, their loci and their relative strengths.  
Note that in principle, one could integrate  the diffusion equation in time and show that 
 secular evolution will drive the distribution function of the underlying disc into 
a state of marginal stability. Such iteration is postponed to another paper
 \citep[][but see section 4.4 for a hint of the expected outcome]{Fouvry2014}. If we rid ourselves of the WKB tepid disc approximation, such equation could also possibly describe the secular diffusion of dark matter cusps in galactic 
centers via the external stochastic feedback processes within the inner baryonic disc. 

One of the limitations of the analysis presented in this paper is the simplified noise model used introduced in equation~\eqref{assumption_noise}, which does not depend on the temporal frequency $\omega$ nor the the radial frequency $k_{r}$ and is therefore only a function of the position in the disc via $J_{\phi}$. Such approximated perturbations intend to recreate the self-induced Poisson shot noise due to the discrete sampling of the smooth distribution function of the disc. A possible improvement would then be to extract the typical power spectrum of the perturbations from numerical experiments, so as to use it as a refined seed for secular diffusion. In a companion paper \citep{FouvryPichonChavanis2014}, we will explore the same WKB limit via the Balescu-Lenard equation. This perspective will enable us to get rid of our crude noise approximation, since it will naturally encompass the \textit{self-induced} shot noise due to finite$-N$ effects and its impact on the secular diffusion of the quasi-stationnary distribution function, as long as we assume that transient tightly wound spirals are governing the evolution of the system. Such an approach will allow us to discuss quantitatively the expected timescales of secular evolution and the respective roles of the drift and diffusion components.
In spite of  this shortcoming, we have shown on this simplified experiment that the secular formalism describes both the 
\textit{nature} of the collisionless system (via its natural frequencies and susceptibility), and its \textit{nurture} via the structure of the external power spectrum.
Hence it provides the natural framework in which to study the cosmic evolution of collisionless systems. 

\vskip 0.5cm
\noindent
\subsection*{Acknowledgements}
Many thanks to  J.~Binney, S.~Prunet and P.~H.~Chavanis for detailed comments.
JBF thanks the {\sc GREAT} program for travel funding and the department of theoretical physics, Oxford for hospitality. 
CP and JBF also thank the Institute of Astronomy, Cambridge, for hospitality 
while this investigation was initiated. 
This work is partially supported by the Spin(e) grants ANR-13-BS05-0005 of the French {\sl Agence Nationale de la Recherche}
and by the ILP LABEX (under reference ANR-10-LABX-63) which  is funded by  ANR-11-IDEX-0004-02.
\bibliographystyle{mn2e}
\bibliography{references}

\appendix

\section{Secular diffusion}
\label{sec:derivation_diffusion_equation}

Let us derive briefly the secular diffusion equation introduced in equation~\eqref{diffusion_equation}. For more details, see the companion paper \cite{FouvryPichonPrunet2014}. We consider a stationary distribution function, $F_{0} (\bm{J})$ (depending only on the actions $\bm{J}$ of the system \citep{Jeans1915}) undergoing an external perturbation. We also suppose that the gravitational potential of the system is fixed so that the action-coordinates mapping ${(\bm{x} , \bm{v}) \mapsto (\bm{\theta} , \bm{J})}$, where $\bm{\theta}$ are the angles canonically associated to the actions, does not depend on time. The distribution function and the Hamiltonian of the system take the form
\begin{equation}
\begin{cases}
\displaystyle F (\bm{J}, \bm{\theta},t) = F_0(\bm{J},t) + f(\bm{J}, \bm{\theta},t)\,,
\\
\displaystyle H(\bm{J}, \bm{\theta}, t) = H_0(\bm{J}) + \psi^e(\bm{J},\bm{\theta},t) + \psi^s(\bm{J},\bm{\theta},t) \, ,
\end{cases}
\label{perturbation_writing}
\end{equation}
where $f$ is the perturbation of the distribution function, $\psi^e$ is the perturbing \textit{exterior} potential generated by an external system, and $\psi^s$ is the \textit{self-response} from the galactic disc generated via self-gravity. We assume that the perturbations are small so that ${f \ll F_{0}}$ and ${\psi^{e}, \psi^{s} \ll \psi_{0}}$, where $\psi_{0}$ is the stationary background potential. We denote by $\bm{\Omega} = {\partial H_0}/{\partial \bm{J}}$ the intrinsic frequencies associated to the actions. Assuming that the disc evolves according to the collisionless Boltzmann equation,
 \begin{equation}
\frac{\mathrm{d} F}{\mathrm{d} t} =\frac{\partial F}{\partial t} +\{H,F\}= 0\,,
 \end{equation}
with $\{\,\,,\,\,\}$ a Poisson bracket,  one obtains from equation~\eqref{perturbation_writing}, in the quasi-linear limit, two equations corresponding to the two timescales of the problem
\begin{equation}
\begin{cases}
\displaystyle \frac{\partial f}{\partial t} + \bm{\Omega} \cdot \frac{\partial f}{\partial \bm{\theta}} - \frac{\partial F_0}{\partial \bm{J}} \cdot \frac{\partial ( \psi^e + \psi^s)}{\partial \bm{\theta}} = 0\,,
\\ \\ 
\displaystyle \frac{\partial F_0}{\partial t} = \frac{1}{(2 \pi)^{2}} \frac{\partial }{\partial \bm{J}} \!\cdot\! \left[ \int \! \mathrm{d}^{2} \bm{\theta} \, f \, \frac{\partial \left[ \psi^{e} + \psi^{s} \right]}{\partial \bm{\theta}}\right] \, .
\end{cases}
\label{evolution_equations}
\end{equation}
The first equation in \eqref{evolution_equations} describes the evolution of the perturbative distribution function $f$ on the fast fluctuating timescale, whereas the second equation describes the long-term evolution of the stationary distribution function $F_{0}$ in action-space. From the first equation, one obtains the expression of the diffusion coefficients appearing in equation~\eqref{diffusion_equation}. We note by ${\bm{m} = (m_{r} , m_{\phi})}$ the Fourier coefficient associated to the Fourier transform with respect to the $2\pi-$periodic angles $\bm{\theta}$, so that the first equation of~\eqref{evolution_equations} takes the form
\begin{equation}
\frac{\partial f_{\bm{m}}}{\partial t} + i \, \bm{m} \!\cdot\! \bm{\Omega} \, f_{\bm{m}} - i \, \bm{m} \!\cdot\! \frac{\partial F_0}{\partial \bm{J}} \,\left[ \psi_{\bm{m}}^e \!+\! \psi_{\bm{m}}^s \right] = 0 \, .
\label{evolution_equation_Fourier_angle}
\end{equation}
Following the matrix method \citep{Kalnajs2}, we introduce a biorthonormal basis of potential $\psi^{(p)}$ and densities $\rho^{(p)}$ satisfying
\begin{equation}
\begin{cases}
\displaystyle \nabla ^{2} \psi^{(p)} = 4 \pi G \rho^{(p)}\,,
\\
\displaystyle \int \!\! \mathrm{d}^3 \bm{x} \, [{\psi^{(p)}} (\bm{x})]^{*} \, \rho^{(q)} (\bm{x}) = - \delta_{p}^{q} \, .
\end{cases}
\label{definition_biorthogonality}
\end{equation}
Given this basis, the exterior potential $\psi^{e}$ and the self-potential $\psi^{s}$ may be written under the form
\begin{equation}
\begin{cases}
\displaystyle \psi^{s} (\bm{x}, t) = \sum_{p}{a_{p} (t) \, \psi^{(p)} (\bm{x})}\,,
\\
\displaystyle \psi^{e} (\bm{x}, t) = \sum_{p}{b_{p} (t) \, \psi^{(p)}(\bm{x})} \, .
\end{cases}
\label{temporal_coefficients}
\end{equation}
To shorten the notations, we also define the associated vectors ${\bm{a} (t) = (a_{1} (t) ,..,a_{p} (t),..)}$ and ${\bm{b} (t) = (b_{1} (t) , ... , b_{p} (t) , ...)}$. The next steps are to solve the equation~\eqref{evolution_equation_Fourier_angle} for $f_{\bm{m}}$, then use the fact that the self-perturbing surface density $\Sigma^{s}$ verifies that ${\Sigma^{s} (\bm{x}) = \int \mathrm{d} \bm{v} \, f (\bm{x},\bm{v})}$ and recall that the transformation ${(\bm{x} , \bm{v}) \mapsto (\bm{\theta} , \bm{J})}$ is canonical so that it satisfies ${\mathrm{d} ^{2} \bm{x} \, \mathrm{d}^{2} \bm{v} = \mathrm{d}^{2} \bm{\theta} \, \mathrm{d}^{2} \bm{J}}$. Given these remarks and assuming that ${\partial F_0 / \partial \bm{J} = cst.}$ on the short timescale, one can show that the equation~\eqref{evolution_equation_Fourier_angle} can be rewritten under the form
\begin{equation}
\widehat{\bm{a}} (\omega) = \widehat{\mathbf{M}}(\omega) \!\cdot\! \left[ \widehat{\bm{a}} (\omega) + \widehat{\bm{b}} (\omega) \right] \, ,
\label{Laplace_a}
\end{equation}
where the response matrix is given by equation~\eqref{Fourier_M}. This expression describes how the perturbations get amplified on the short timescale.

 The next step is to use the second equation of~\eqref{evolution_equations} to capture the secular evolution of the quasi-stationary distribution function $F_{0}$ in action-space. Introducing the sum of the two gravitational perturbations ${\bm{c} (t) = \bm{a} (t) \!+\! \bm{b} (t)}$, one can show that the second equation of~\eqref{evolution_equations} takes the form
\begin{equation} \!\!\!\!
\frac{\partial F_0}{\partial t} = \sum_{\bm{m}} \bm{m} \!\cdot\! \frac{\partial}{\partial \bm{J}} \left[ D_{\bm{m}} (\bm{J},t) \, \bm{m} \!\cdot\! \frac{\partial F_0}{\partial \bm{J}} \right] \, ,
\label{secular_timeaveraged_equation}
\end{equation}
where the diffusion coefficients ${D_{\bm{m}} (\bm{J})}$ are given by
\begin{equation}
D_{\bm{m}} (\bm{J} , t) \!=\! \sum_{p , q} \psi_{\bm{m}}^{p} \psi_{\bm{m}}^{(q) *} c_{q}^{*} (t) \!\!\int_{- \infty}^{t} \!\!\!\! \mathrm{d} \tau \, e^{- i \bm{m} \cdot \bm{\Omega} (t - \tau)} c_{p} (\tau) \, .
\label{expression_Dm_c}
\end{equation}
Using the matrix relation~\eqref{Laplace_a}, one can note that ${\widehat{\bm{c}} = [ \mathbf{I} \!-\! \widehat{\mathbf{M}} ]^{-1} \!\!\cdot\! \, \widehat{\bm{b}}}$, so that equation~\eqref{expression_Dm_c} can be expressed only as a function of the external perturbation $\widehat{\bm{b}}$. The final step to derive the expression~\eqref{expression_diffusion_coefficients_all} of the diffusion coefficients is to introduce statistical averages, in order to consider only the mean response of a typical disc. We introduce the operation of ensemble average over many realizations of external perturbations, noted as $\left< \,.\, \right>$. When taking the ensemble average of equation~\eqref{secular_timeaveraged_equation}, one can assume that the response matrix $\widehat{\mathbf{M}}$, the distribution function $F_{0}$ and its gradient $\partial F_{0} / \partial \bm{J}$ do not change significantly from one realization to another. Indeed, we intend to describe the effect of an averaged fluctuation on a given distribution function $F_{0}$ representing a mean disc. Under these assumptions, equation \eqref{secular_timeaveraged_equation} becomes
\begin{equation}
\begin{aligned}
\frac{\partial F_0}{\partial t} = \sum_{\bm{m}} \bm{m} \!\cdot\! \frac{\partial}{\partial \bm{J}} \left[ \!\bigg< D_{\bm{m}} (\bm{J} , t) \!\bigg> \, \bm{m} \!\cdot\! \frac{\partial F_0}{\partial \bm{J}} \right] \, .
\end{aligned}
\label{secular_doubleaveraged_equation}
\end{equation}
We finally introduce a stationarity hypothesis for the time evolution of the exterior perturbation on short timescales and therefore define the temporal auto-correlation of the exterior perturbation as
\begin{equation}
\mathbf{C}_{k l} (t_{1} - t_{2}) =  \left< b_{k} (t_{1}) \, b_{l}^{*} (t_{2}) \right> \, .
\label{definition_autocorrelation_exterior}
\end{equation}
From this definition, one can compute the ensemble average of the Fourier transformed term $\big< \widehat{b}_{k} \, \widehat{b}_{l}^{\, *}  \big>$, to obtain
\begin{equation}
\left< \widehat{b}_{k} (\omega) \, \widehat{b}_{l}^{*} (\omega ') \right> = 2 \pi \, \delta_{\rm D} (\omega \!-\! \omega ') \, \widehat{\mathbf{C}}_{k l} (\omega) \, ,
\label{autocorrelation_exterior_Fourier}
\end{equation}
where $\widehat{\mathbf{C}}$ stands for the temporal Fourier transform of the autocorrelation function of the exterior potential. Injecting equation~\eqref{autocorrelation_exterior_Fourier} into equation~\eqref{secular_doubleaveraged_equation} and using hermiticity arguments to show that only the real part of the diffusion coefficients matters, one obtains the final writing of the secular diffusion equation, which reads
\begin{equation}
\begin{aligned}
\frac{\partial F_0}{\partial t} = & \sum_{\bm{m}} \bm{m} \!\cdot\! \frac{\partial}{\partial \bm{J}} \bigg[  \bm{m} \!\cdot\! \frac{\partial F_0}{\partial \bm{J}} \sum_{p , q} \frac{1}{2} \, \psi^{(p)}_{\bm{m}} \psi^{(q) *}_{\bm{m}} 
\\
& \left[ [ \mathbf{I} \!-\! \widehat{\mathbf{M}}]^{-1} \!\cdot \widehat{\mathbf{C}} \cdot  [\mathbf{I} \!-\! \widehat{\mathbf{M}}]^{-1} \right]_{p q} \!( \bm{m} \!\cdot\! \bm{\Omega})\,  \bigg] \, ,
\end{aligned}
\label{secular_doubleaveraged_equation_final}
\end{equation}
which is the same expression of the diffusion equation as introduced in equations~\eqref{diffusion_equation} and~\eqref{expression_diffusion_coefficients_all}.

\section{WKB  coefficients}
\label{sec:WKB_approximation}

Let us derive briefly in this appendix the diffusion coefficients $D_{\bm{m}} (\bm{J})$ in the WKB limit. The key ingredient  is to introduce specific basis elements well-suited to represent tightly wound spirals. We consider potential elements given by
\begin{equation}
\psi^{[ k_{\phi}, k_{r} , R_{0}]} (R,\phi) \!=\! \frac{\mathcal{A} \, e^{ i ( k_{\phi} \phi + k_{r} R) }}{(\pi \sigma^{2})^{1\!/\!4}} \, \exp \!\left[ - \frac{(R \!-\! R_{0})^{2}}{2 \sigma^2} \right] \, ,
\label{definition_basis_WKB}
\end{equation}
where the basis elements are indexed by three quantities. Here $R_0$ is the central radius around which the Gaussian is centered, $k_{\phi}$ is an azimuthal number representing the angular component of the basis elements, $k_r$ corresponds to the radial frequency of the potential, and ${\mathcal{A} = \sqrt{{G}/ {|k_r| R_0} }}$ is an amplitude tuned in order to ensure the correct normalization of the basis as imposed by equation~\eqref{definition_biorthogonality}. Finally, $\sigma$ is a scale-separation parameter ensuring the biorthogonality of the basis elements. Introducing the typical radial size of the system $R_{\rm sys}$, one can show that under the assumptions of tight-winding ${k_{r} R_{0} \gg 1}$ and of scale-separation ${k_r \sigma \gg {R_{\rm sys}}/{\sigma}}$, the associated surface density elements $\Sigma^{[ k_{\phi} , k_{r} , R_{0}]}$, obtained via Poisson's equation, are given by
\begin{equation}
\Sigma^{[ k_{\phi}, k_{r} , R_{0}]} (R, \phi) = - \frac{| k_{r} |}{2  \pi G} \, \psi^{[ k_{\phi}, k_{r} , R_{0}]} (R, \phi) \, .
\label{definition_Sigma_WKB}
\end{equation}
In order to ensure the biorthogonality of the basis elements, one can show that the \textit{distance} between two basis elements $\psi^{1}$ and $\psi^{2}$, represented by ${ \Delta R_{0} = R_{0}^{1} \!-\! R_{0}^{2} }$ and ${ \Delta k_{r} = k_{r}^{1} \!-\! k_{r}^{2} }$ must satisfy
\begin{equation}
\begin{cases}
\begin{aligned}
\displaystyle \Delta R_{0} &\gg \sigma &\text{or}&\;\;\; \Delta R_{0} = 0 \, ,
\\
\displaystyle \Delta k_{r} &\gg \frac{1}{\sigma} &\text{or}&\;\;\; \Delta k_{r} = 0\, .
\end{aligned}
\end{cases}
\label{separation_assumption}
\end{equation}
With such explicit basis elements, one can compute their Fourier transforms with respect to the actions which read
\begin{equation}
\begin{aligned}
\psi^{[k_{\phi}, k_{r} , R_0]}_{\bm{m}} (\bm{J}) = \;& \delta^{k_{\phi}}_{m_{\phi}} \, e^{i m_{r} \theta^{0}_{R}} \, \frac{\mathcal{A}}{(\pi \sigma^{2})^{1\!/\!4}} \, e^{i k_{r} R_{g} }
\\
& \mathcal{J}_{m_{r}} (H_{k_{r}}) \, \exp \!\left[\! - \frac{(R \!-\! R_{0})^{2}}{2 \sigma^2} \!\right] \, ,
\end{aligned}
\label{Fourier_basis_WKB}
\end{equation}
where $\mathcal{J}_{m_{r}}$ is the Bessel function of the first kind of index $m_{r}$. Thanks to the WKB approximation which assumes that $k_{r} R_{g} \gg 1$, the amplitude $H_{k_{r}}$ and the phase shift $\theta^{0}_{R}$ are given by
\begin{equation}
 H_{k_{r}} \simeq \sqrt{\tfrac{\displaystyle 2 J_{r}}{\displaystyle \kappa (J_{\phi})}} \, k_{r} \;\;;\;\; \theta^{0}_{R} \simeq - \pi / 2 \, .
\label{expression_H_theta_0}
\end{equation}
Within this framework, one can now evaluate the response matrix elements from equation~\eqref{Fourier_M}. Because of the assumptions of tight winding, one can show that the response matrix is diagonal, so that two distinct WKB basis elements cannot interact one with another. Finally, we introduce the additional assumption that the galactic disc considered is tepid, so that ${ \big| \partial F_0 / \partial J_r \big| \gg \big| \partial F_{0} / \partial J_{\phi} \big| }$. Keeping only the dominant terms, the expression of the diagonal response matrix eigenvalues for a tepid disc reads
\begin{equation} \hskip -0.25cm
\widehat{\mathbf{M}}_{\left[ k_{\phi}^{p} , k_{r}^{p} ,R_0 \right] , \left[ k_{\phi}^{q} , k_{r}^{q} , R_0 \right]} = \delta_{k_{\phi}^{p}}^{k_{\phi}^{q}} \, \delta_{k_{r}^{p}}^{k_{r}^{q}} \, \frac{2 \pi G \, \xi \, \Sigma \, | k_r |}{\kappa^2 (1 \!-\! s^2)} \, \mathcal{F} (s, \chi) \, ,
\label{expression_M_diagonal_tepid}
\end{equation}
which is consistent with equation~\eqref{expression_eigenvalues}. Using the fact that response matrix is now diagonal, the expression~\eqref{expression_diffusion_coefficients_all} of the diffusion coefficients is easier to compute.

The last step of the derivation is to express the basis coefficients $b_{p}$ as a function of the perturbing exterior potential $\psi^{\rm ext}$ and to replace the sum on the basis index $k_{r}$ and $R_{0}$ by continuous integrals, using Riemann formula ${\sum f (x) \Delta x \simeq \int \!\mathrm{d} x  f (x)}$. As we have assumed that the exterior perturbation had the simple dependence from equation~\eqref{assumption_noise}, one finally obtains the expression of the diffusion coefficients in the WKB limit given in equation~\eqref{diffusion_coefficients_WKB}.

\balance

\section{Growth rate of ridge}
\label{sec:ridge_growthrate}

Let us estimate the rate at which the ridge in action space observed in S12 develops. As the evolution of the system is dominated by the ILR resonance, we introduce the associated fast and slow actions from equation~\eqref{definition_fast_actions}. In order to shorten the notations, we note the slow and fast actions as $J_{\rm ILR}^{s} = J_{s}$ and $J_{\rm ILR}^{f} = J_{f}$. When considering only one resonance, the diffusion equation~\eqref{diffusion_equation_divergence} becomes mono-dimensional and, for a given value of $J_{f}$, up to constant prefactor in $|\bm{m}|$, takes the form
\begin{equation}
\frac{\partial F_{0}}{\partial t} = \frac{\partial }{\partial J_{s}} \left[ D (J_{s}) \, \frac{\partial F_{0}}{\partial J_{s}} \right] \, .
\label{diffusion_equation_unidimensional}
\end{equation}
Equation~\eqref{diffusion_equation_unidimensional} corresponds to a $1D$ inhomogeneous heat equation. We suppose that initially the distribution function is concentrated within a narrow region in $J_{s}$, and we use a method based on self-similar solutions in order to estimate the typical growth rate of the ridge. Therefore, we introduce a self-similar Ansatz for the shape of the $1D$ distribution function of the form (see formula $1.3.6.8$ from~\cite{HandbookLinearPartial} and eq.~${(1.70)}$ from~\cite{BinneySecular2013})
\begin{equation}
F (J_{s} , t) = \frac{1}{t^{a}} \, F_{\rm sc} \!\left[ \frac{J_{s}}{t^{a}}\right] \, .
\label{ansatz_self_similar}
\end{equation}
The coefficient $a>0$ is for the moment unconstrained but will encode the speed with which the scatter of the distribution function increases. Indeed, one immediately obtains the time dependence of the standard deviation of the distribution function in the $J_{s}-$direction as
\begin{equation}
\sigma_{J_{s}} = \sqrt{ \left< J_{s}^{2} \right>} \propto t^{a} \, .
\label{scatter_growth_rate}
\end{equation}
In order to be able to draw qualitative conclusions from $a$, we will assume that the anistropic diffusion coefficients satisfy a scaling property of the form
\begin{equation}
D (k \, J_{s}) = k^{b} \, D (J_{s}) \, ,
\label{scaling_diffusion_coefficients}
\end{equation}
for any ${k > 0}$ and where ${0 \leq b < 2}$ captures the structure of the anistropic diffusion. The homogeneous heat equation corresponds to the case ${D(J_{s}) = cst}$, so that one has $b = 0$. Starting from the ansatz~\eqref{ansatz_self_similar} and using the rescaling change of variables ${K = J_{s} / t^{a}}$, one can show that the diffusion equation~\eqref{diffusion_equation_unidimensional} becomes
\begin{equation}
- \frac{a}{t^{a + 1}} \left[ F_{\rm sc} \!\left[ K \right] + K \frac{\partial F_{\rm sc}}{\partial K} \right] = t^{a b - 3 a} \frac{\partial }{\partial K} \left[ D (K) \frac{\partial F_{\rm sc}}{\partial K} \right] \, . \nonumber
\label{full_differential_equation_scale_invariant}
\end{equation}
In order to have an equation valid for all time, both side must necessarily have the same time dependence, so that we obtain the condition
\begin{equation}
ab = 2 a \!-\! 1 \, .
\label{relation_a_b_scaling}
\end{equation}
As a consequence, given  the scaling coefficient $b$, one constrains $a$ and then predicts the rate with which the standard deviation will increase during the diffusion. For example, for the homogeneous heat equation, one has $b = 0$, so that $a = 1/2$. Using equation~\eqref{scatter_growth_rate}, one obtains that $\sigma_{J_{f}} \propto \sqrt{t}$, which is the usual growth rate of the scattering expected for an homogeneous diffusion. Moreover, one can also obtain the shape of the scale-invariant distribution function $F_{\rm scal}$ through the ordinary differential equation
\begin{equation}
\frac{\partial }{\partial K} \left[ D (K) \frac{\partial F_{\rm sc}}{\partial K} \right]  + a \left[ F_{\rm sc} \!\left[ K \right] + K \frac{\partial F_{\rm sc}}{\partial K} \right] = 0 \, .
\label{equation_F_sc}
\end{equation}
Using an ansatz of the form $F_{\rm scal} (K) = \exp [- K^{\alpha} / \beta]$ and assuming that $D (J_{s}) = D_{0} \, [J_{s}]^{b}$, the solutions of the ordinary differential equation~\eqref{equation_F_sc} take the form
\begin{equation}
F_{\rm scal} (K) \propto \exp \left[ - \frac{K^{2 - b}}{D_{0} (2 \!-\! b)^{2} } \right] \, .
\label{general_solution_scale_invariant}
\end{equation}
In the case of the homogeneous diffusion $b = 0$, one obtains as expected a scale-invariant solution given by a Gaussian. We may now restrict ourselves to the case where the ILR resonace dominates, for which we want to estimate the temporal growth rate of an initially dense spot located in $J_{\phi}^{0}$, on the $J_{r} = 0$ axis. Using the change of variables to the fast and slow actions, we introduce $J_{0}^{s}$ and $J_{0}^{f}$ such that ${(J_{\phi}^{0}, J_{r} = 0) \to (J_{0}^{s} , J_{0}^{f})}$. As the diffusion takes place only in the $J_{s}-$direction, we need to study the variations of the diffusion coefficients given by the function ${ j_{s} \mapsto D_{\rm ILR} (J_{0}^{s} \!+\! j_{s} , J_{0}^{f} ) }$. We start from the expression of the diffusion coefficients obtained in equation~\eqref{diffusion_coefficients_small_denominators}. As the diffusion starts from $J_{r} = 0$, we may perform of limited development of the Bessel function at the origin, recalling that for ${ x \ll 1 }$, ${ \mathcal{J}_{m_{r}} (x) \propto x^{|m_{r}|} }$, so that we obtain a dependence of the form
\begin{equation}
D_{\rm ILR} (J_{0}^{s} + j_{s} , J_{0}^{f} ) \propto j_{s} \left[ \frac{ (\Delta k_{\lambda})^{2} k_{\rm max}^{2} \, \Sigma_{\rm t} }{ \kappa \, (1 \!-\! \lambda_{\rm max} )^{2} } \right] (J_{\phi} [J_{0}^{s} \!+\! j_{s} , J_{0}^{f}] ) \, , \nonumber
\label{calculation_scaling_fast_actions}
\end{equation}
where the term between brackets has to be evaluated at the angular momentum $J_{\phi}$ corresponding to the fast and slow actions coordinates ${ (J_{0}^{s} \!+\! j_{s} , J_{0}^{f}) }$. This term tends a finite non-zero value for ${ j_{s} \to 0 }$. As a consequence, for ${j_{s} \to 0}$, one finally obtains the shortened dependence
\begin{equation}
D_{\bm{m}} (J_{0}^{s} \!+\! j_{s} , J_{0}^{f} ) \propto j_{s} \, .
\label{asymptotic_scaling_fast_actions}
\end{equation}
One can immediately conclude that for an ILR-dominated diffusion along its associated slow action, the scaling coefficient of the anisotropic diffusion coefficients $D_{\rm ILR}$ is given by ${ b_{\rm ILR} = 1 }$. Using the relation~\eqref{relation_a_b_scaling}, one obtains that $a_{\rm ILR} = 1$, so that the temporal growth rate of the standard deviation along the $J_{\rm ILR}^{s}-$direction is given by
\begin{equation}
\sigma_{J_{\rm ILR}^{s}} \propto t \, .
\label{temporal_growth_rate_ILR}
\end{equation}
This is the scaling presented in the main text.

\vfill
\eject

\label{lastpage}
\end{document}